\documentclass[amsmath,amssymb,aps,prl,reprint,floatfix]{revtex4-2}

\usepackage{graphicx}% Include figure files
\usepackage{physics}
\usepackage{dcolumn}% Align table columns on decimal point
\usepackage{bm}% bold math
\usepackage{lipsum, babel}

\usepackage{comment}
\usepackage[svgnames]{xcolor}
\usepackage{xstring}
\newcommand{\EatOneArg}[1]{}

\newcommand{\addMS}[1]{\textcolor{black}{#1}}
\newcommand{\addSAS}[1]{\textcolor{black}{#1}}

\usepackage{amsmath}
\usepackage{graphicx}% Include figure files
\usepackage{dcolumn}% Align table columns on decimal point
\usepackage{bm}% bold math
\usepackage[hidelinks]{hyperref}% add hypertext capabilities

\usepackage{xcolor}

\newcommand{\diff}{\mathrm{d}}

\newcommand{\oma}{\omega}

\newcommand{\cg}{c_g}

\newcommand{\ea}{e_\alpha}
\newcommand{\eb}{e_\beta}
\newcommand{\ec}{e_\gamma}

\renewcommand{\oma}{\omega_\alpha}
\newcommand{\omb}{\omega_\beta}
\newcommand{\omg}{\omega_\gamma}

\newcommand{\mean}[1]{\overline{#1}}

\newcommand{\leri}[1]{\left(#1\right)}

\newcommand{\obs}[1]{\langle #1 \rangle }
\usepackage[nolist]{acronym}
\usepackage{xcolor}
\usepackage{xspace}
\newcommand{\shahaf}[1]{{\color{purple} [Shahaf: #1]\xspace}}
\usepackage{color}
\definecolor{lightblue}{cmyk}{1,0.3,0,0.2}
\definecolor{purplePink}{cmyk}{0,1,0,0.2}
\hypersetup{colorlinks, bookmarksnumbered, citecolor=lightblue, linkcolor=lightblue, pdfstartview=FitH, urlcolor=purplePink, linktocpage}

\begin{document}

%lyx shortcuts
\global\long\def\dyad#1{\underline{\underline{\boldsymbol{#1}}}}%
\global\long\def\ubar#1{\underbar{\ensuremath{\boldsymbol{#1}}}}%
\global\long\def\integer{\mathbb{Z}}%
\global\long\def\natural{\mathbb{N}}%
\global\long\def\real#1{\mathbb{R}^{#1}}%
\global\long\def\complex#1{\mathbb{C}^{#1}}%
\global\long\def\defined{\triangleq}%
\global\long\def\trace{\text{trace}}%
\global\long\def\del{\nabla}%
\global\long\def\cross{\times}%
\global\long\def\diff#1#2{\frac{\partial#1}{\partial#2}}%
\global\long\def\Diff#1#2{\frac{d#1}{d#2}}%
\global\long\def\bra#1{\left\langle #1\right|}%
\global\long\def\ket#1{\left|#1\right\rangle }%
\global\long\def\braket#1#2{\left\langle #1|#2\right\rangle }%
\global\long\def\ketbra#1#2{\left|#1\right\rangle \left\langle #2\right|}%
\global\long\def\identity{\mathbf{1}}%
\global\long\def\paulix{\begin{pmatrix}  &  1\\
 1 
\end{pmatrix}}%
\global\long\def\pauliy{\begin{pmatrix}  &  -i\\
 i 
\end{pmatrix}}%
\global\long\def\pauliz{\begin{pmatrix}1\\
  &  -1 
\end{pmatrix}}%
\global\long\def\sinc{\mbox{sinc}}%
\global\long\def\four{\mathcal{F}}%
\global\long\def\dag{^{\dagger}}%
\global\long\def\norm#1{\left\Vert #1\right\Vert }%
\global\long\def\hamil{\mathcal{H}}%
\global\long\def\tens{\otimes}%
\global\long\def\ord#1{\mathcal{O}\left(#1\right)}%
\global\long\def\undercom#1#2{\underset{_{#2}}{\underbrace{#1}}}%
 
\global\long\def\conv#1#2{\underset{_{#1\rightarrow#2}}{\longrightarrow}}%
\global\long\def\tg{^{\prime}}%
\global\long\def\ttg{^{\prime\prime}}%
\global\long\def\clop#1{\left[#1\right)}%
\global\long\def\opcl#1{\left(#1\right]}%
\global\long\def\broket#1#2#3{\bra{#1}#2\ket{#3}}%
\global\long\def\div{\del\cdot}%
\global\long\def\rot{\del\cross}%
\global\long\def\up{\uparrow}%
\global\long\def\down{\downarrow}%
\global\long\def\Tr{\mbox{Tr}}%

\global\long\def\per{\mbox{}}%
\global\long\def\pd{\mbox{}}%
\global\long\def\p{\mbox{}}%
\global\long\def\ad{\mbox{}}%
\global\long\def\a{\mbox{}}%
\global\long\def\la{\mbox{\ensuremath{\mathcal{L}}}}%
\global\long\def\cm{\mathcal{M}}%
\global\long\def\cg{\mbox{\ensuremath{\mathcal{G}}}}%
%end lyx shortcuts

\preprint{APS/123-QED}

\title{Sign-Indefinite Helicity and the Structure of Weak Turbulence in Inertial and Non-Hermitian Waves}
\author{Shahaf Aharony Shapira$^{\dagger}$ and Michal Shavit$^{*}$\\
\emph{\normalsize{}$^{\dagger}$ Department of Physics of Complex Systems, Weizmann Institute of Science, Rehovot 7610001, Israel}\\
\emph{\normalsize{}$^{*}$ Courant Institute of Mathematical Sciences, New
York University, NY 10012, USA.}}
% \affiliation{Department of Physics of Complex Systems, Weizmann Institute of Science, Rehovot 7610001, Israel}

\begin{abstract}
%\addMS{The most important thing to add is that sign-definite helicity interactions transfer energy backward also in the sign-indefinite case!}

We investigate how sign-indefinite quadratic invariants shape turbulent cascades in incompressible flows with broken time-reversal symmetry, where the dynamics support strongly anisotropic dispersive waves. Focusing on rotating Euler flow and odd-viscous Euler flow, we isolate the wave component and study the corresponding weak-turbulence kinetic equation. We show that helicity conservation substantially simplifies the kinetic equation. Fixing the energy flux via a natural gauge choice, we identify the turbulent spectrum as the unique scale-invariant solution that sustains a constant energy flux from large to small scales. Under a mild approximation motivated by the accumulation of energy near slow modes, we compute the leading angular dependence and uncover an integrable singularity along the slow-mode curve, that agrees with previous results.
We then demonstrate that helicity reorganizes cascade directions at the level of resonant triads. Although helicity is globally sign-indefinite, the helical decomposition splits it into sign-definite contributions on each polarization branch. Triads whose three legs lie on the same branch behave as if constrained by a sign-definite invariant and drive an upscale transfer of energy, producing systematic backscatter even when the net cascade is direct. In the helicity-definite limit (single-branch dynamics), the kinetic equation admits an additional scale-invariant solution associated with helicity transport. 
Finally, we validate the analytical predictions by numerically evaluating the collision integral in the strongly anisotropic limit, revealing a family of stationary solutions in that regime.

\end{abstract}

\maketitle
% List of acronyms
\begin{acronym}
    \acro{3D}{three-dimensional}
    \acro{RHS}{right-hand side}
    \acro{w.r.t.}{with respcet to}
    \end{acronym}
%\tableofcontents
%\section{\label{sec:level1}Introduction}

\textbf{\textit{Introduction}} 
A central question in turbulence is how conservation laws constrain the direction of energy transfer across scales.
Understanding how quadratic invariants shape turbulent cascades has been a central theme of fluid dynamics since the pioneering works of Kolmogorov and Kraichnan \cite{kolmogorov1991local, kraichnan1967inertial}. In isotropic \ac{3D} \addSAS{hydrodynamic} turbulence, energy cascades directly toward small scales, while helicity, being sign-indefinite, does not by itself enforce an inverse cascade. By contrast, in systems where the second quadratic invariant is sign-definite—such as enstrophy in two-dimensional turbulence—energy transfer is strongly constrained and an inverse cascade emerges: Fjørtoft’s classical argument shows that the simultaneous conservation of energy and enstrophy forces energy to migrate to larger scales, while enstrophy flows to smaller ones \cite{fjor1953changes}. More broadly, in weakly nonlinear wave systems the same principle leads to inverse cascades associated with conserved invariants, captured by the Kolmogorov–Zakharov spectra of wave turbulence theory\addSAS{\cite{zakharov1992kolmogorov,galtier2022physics}}. These examples suggest that even invariants that are globally sign-indefinite may nonetheless influence cascade directions when the dynamics organizes interactions into subsets where they become effectively sign-definite.

\begin{figure}[h] 
\includegraphics[scale=0.3]{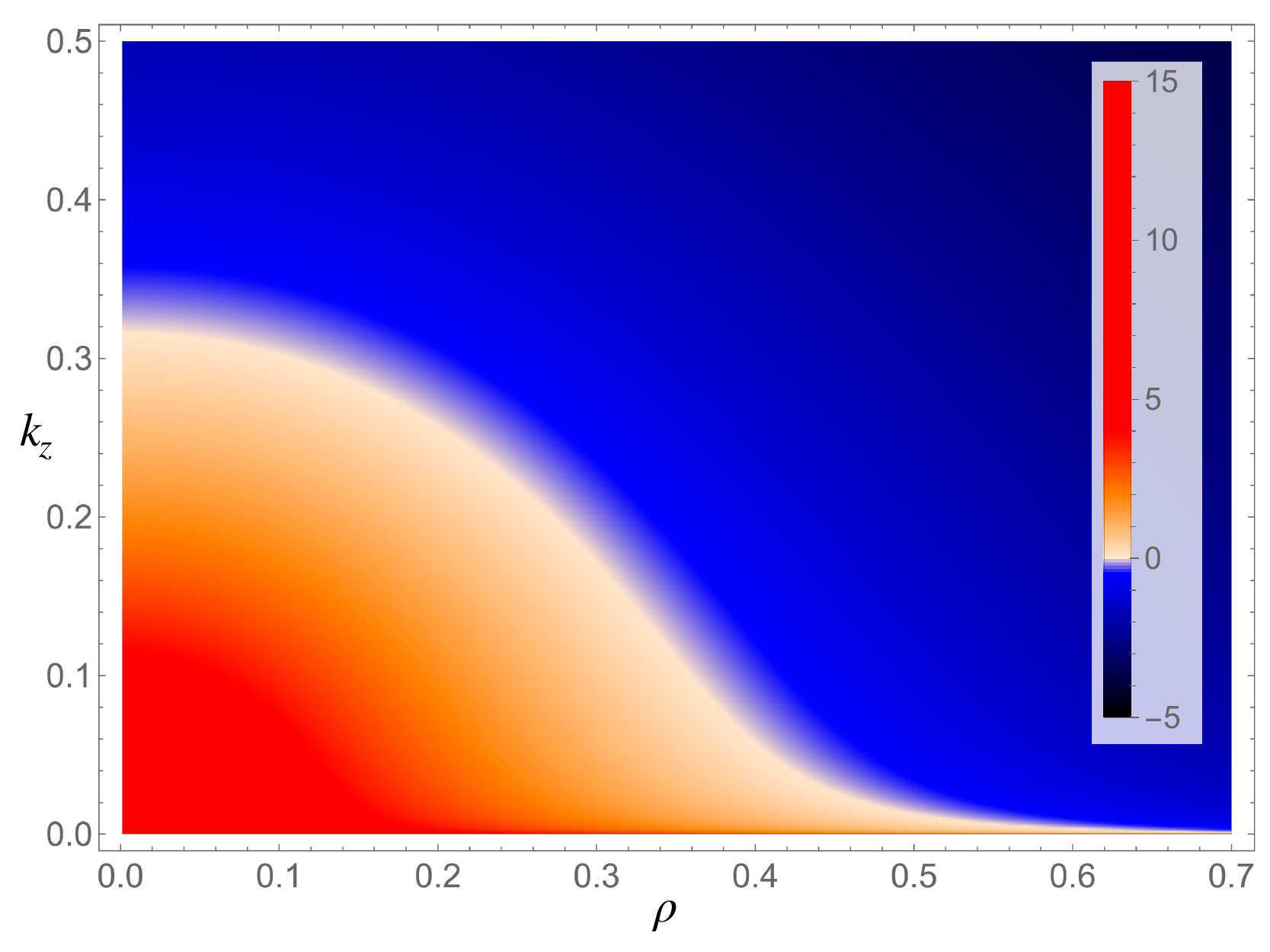}
\caption{\label{fig:2dspec} A heatmap of the logarithm of the energy spectrum $e_k=C_0k^{-3}|\cos\theta_k|^{-1/2}$ \addSAS{versus the wavenumber $\bm{k}$, in cylindrical coordinates: $\rho \equiv \sqrt{k_x^2+k_y^2}$, and $k_z$}. This corresponds to the energy spectrum outside an angular source centered at radius $k_f$ and emitting a radial energy flux $\Pi^r(\theta_k)=\Pi_0 |\cos\theta_k|^{-1} \hat{k}/ k$.}
\end{figure}

A natural setting in which such questions arise is turbulence in systems where time-reversal symmetry is broken and the dynamics supports chiral dispersive waves. Two paradigmatic examples are rotating fluids \cite{poincare1885equilibre,greenspan1969theory, galtier2003weak}, where the Coriolis force singles out the axis of rotation, and fluids with odd (Hall) viscosity, where parity and time-reversal symmetry are simultaneously broken by the stress tensor \cite{avron1998odd, de2024pattern, fruchart2023odd, de2025non}. In both cases, the linearized dynamics support anisotropic dispersive waves: inertial waves in rotating flows and ``odd waves'' in odd-viscous fluids. The presence of such waves naturally leads to a weak-turbulence regime governed by (pseudo-)resonant wave interactions, which we can study analytically. %While weak turbulence of inertial waves in rotating fluids has been extensively developed \cite{galtier2003weak}, the corresponding interacting-wave theory for odd-viscous fluids has only recently begun to be explored \cite{}. Both were studied in a very anisotropic approximation, our study is 

These systems therefore provide a controlled framework for investigating how helicity influences turbulent cascades in anisotropic, wave-dominated flows. In the helical basis, helicity decomposes into sign-definite contributions associated with the two circular polarizations of each Fourier mode. This decomposition allows one to isolate subsets of resonant interactions in which helicity becomes effectively positive-definite. As we show below, interactions within such subsets reverse the direction of energy transfer: sign-definite helicity interactions drive an inverse cascade, even though the full system admits a direct cascade. This complements earlier observations that small-scale intermittency is reduced in systems with such constraints compared to fully isotropic \ac{3D} turbulence \cite{chen2024odd}. \addMS{Related phenomena were previously observed in the context of strong turbulence by \cite{biferale2012inverse}, where the 3D Navier--Stokes equations were projected onto positive-helicity states. In that setting, both energy and helicity become sign-definite invariants, leading to an inverse energy cascade analogous to the role played by enstrophy conservation in 2D turbulence. In contrast, the present work considers weak turbulence in the full system, where both helicity signs remain active. Nevertheless, we show analytically that resonant interactions restricted to modes of the same helicity sign still transfer a fraction of the energy upscale.}

%These systems therefore provide a controlled framework for investigating how helicity influences turbulent cascades in anisotropic, wave-dominated flows. In the helical basis, helicity decomposes into sign-definite contributions associated with the two circular polarizations of each Fourier mode. This decomposition allows one to isolate subsets of resonant interactions in which helicity becomes effectively positive-definite. As we show below, interactions within such subsets reverse the direction of energy transfer: sign-definite helicity interactions drive an inverse cascade, even though the full system admits a direct cascade. This complements earlier observations that small-scale intermittency is reduced in systems with such constraints compared to fully isotropic \ac{3D} turbulence \cite{chen2024odd}. \addMS{In cite~\cite{biferale2012inverse} }

%In particular, the sign-definiteness of helicity at the level of resonant triads is sufficient to enforce systematic upscale transfer.

To characterize these cascades, we introduce a natural gauge fixing for the energy flux that separates isotropic and anisotropic contributions. This allows us to identify the unique scale-invariant solutions that support a nonzero radial flux. Exploiting the scale invariance of the governing equations, and without imposing the strongly anisotropic limit, we derive the scale-invariant component of the energy spectrum. Our analysis shows that anisotropic wave turbulence fundamentally organizes energy transfer through angular structure on the unit sphere, where forcing may be applied without loss of generality. From this angular distribution, the cascade proceeds radially—though anisotropically—in a scale-invariant manner. This viewpoint improves upon previous treatments that impose strong anisotropy at the outset and consequently produce nonphysical divergences in the predicted energy spectrum.

To determine the angular structure of the spectrum, we treat anisotropy in a weaker sense and compute the asymptotic angular dependence. This reveals weak but integrable singularities near the slow-mode manifold, where wave energy is expected to accumulate, see Figure \ref{fig:2dspec}. We then numerically evaluate the collision integral in the standard strongly anisotropic limit to confirm and illustrate these analytical predictions.

Odd-wave turbulence connects fundamental fluid mechanics with broader themes across physics. Odd viscosity emerges as an effective transport coefficient in active matter and certain quantum fluids, while rotating flows are ubiquitous in geophysical and astrophysical contexts. Both systems illustrate how the breaking of discrete symmetries—parity and time reversal—can qualitatively reorganize cascade phenomenology. In this work, we focus on the wave component of the flow. While our results indicate a concentration of wave energy near slow modes, we do not analyze the slow modes themselves. Their dynamics, particularly in the presence of condensates, have recently been investigated in the context of rotating turbulence \cite{gome2025helicity, gome2025waves}.

%The paper is organized as follows...

\textbf{\textit{Governing equations}}
Consider the Euler equation in \ac{3D} space written in vorticity form
\begin{equation} \label{eq:euler}
\partial_{t}\bm{\omega} = \leri{\bm{\omega}\cdot\nabla}\mathbf{v} - \leri{\mathbf{v}\cdot\nabla}\bm{\omega}\,,
\end{equation}
where $\mathbf{v}$ is the velocity field, $\bm{\omega} = \nabla\times\mathbf{v}$ is the vorticity and incompressibility is assumed, $\nabla\cdot \mathbf{v} = 0$. The Euler equation has two quadratic invariants: the positive-definite kinetic energy, $E = \frac{1}{2} \int |\mathbf{v}|^2\,d\mathbf{x}$, and the (sign-indefinite) helicity,
$H = \int \mathbf{v}\cdot\bm{\omega}\,d\mathbf{x}$. Eq.~\eqref{eq:euler} can be written as a noncanonical Hamiltonian system on the dual of the divergence-free vector fields~\cite{olver1982nonlinear}:
\begin{equation} \label{eq:hamil}
% \partial_{t}\bm{\omega} = \mathcal{J}\,\frac{\delta E}{\delta\bm{\omega}},
\partial_t \bm{\omega} = \mathcal{J} \leri{\frac{\delta E}{\delta \bm{\omega}}}\,,
\end{equation}
%\begin{equation} \label{eq:hamil}
%\partial_{t}\bm{\omega} = \{\bm{\omega},E\}_\mathcal{J} = %\mathcal{J}\,\frac{\delta E}{\delta\bm{\omega}},
%\end{equation}
where the Poisson operator $\mathcal{J}$ is the skew-symmetric form
\begin{equation} \label{eq:poisson}
% \mathcal{J} = \left(\bm{\omega}\cdot\nabla - \nabla\bm{\omega}\right)\,\nabla\times,
\mathcal{J} \leri{\bm{a}} = - \curl{\leri{\bm{\omega} \cross \leri{ \curl \bm{a}}}} \,.
\end{equation}
% and $(\nabla\bm{\omega})_{ij} = \partial_j \omega_i$ denotes the Jacobian matrix of $\bm{\omega}$.
From~\eqref{eq:hamil} it follows that the kinetic energy is conserved due to time-translation invariance. The helicity, which is a \emph{Casimir} of the Poisson structure~\eqref{eq:poisson}, is also conserved because the velocity lies in the kernel of $\mathcal{J}$. See~\cite{SM} for more details. Geometrically, this reflects the Lie advection of vorticity by the flow, which preserves the topology of vortex lines and hence helicity.
\addMS{In the present work, we study how helicity conservation simplifies the kinetic equation and redistributes energy among resonant interactions. However, we do not investigate whether helicity, as a Casimir, may further constrain the turbulent state or influence its statistical properties. This remains an interesting direction for future work, as discussed in the conclusion.}

\textbf{\textit{Helical decomposition.}}
In a periodic box of size $L^{3}$, we expand the velocity in eigenfunctions of the curl operator $\bm{h}_{s_\alpha}$,
\begin{align}
    \mathbf v(\mathbf x,t) =\sum_{\alpha} v_\alpha(t)\, \mathbf h_{s_\alpha}(\mathbf k_\alpha) e^{\,i\mathbf k_\alpha \cdot \mathbf x} \,,
\end{align}
%$\mathbf v(\mathbf x,t)=\sum_{\mathbf k}\sum_{s=\pm 1} v_s(\mathbf k,t)\mathbf h_s(\mathbf k)\,e^{i\mathbf k\cdot \mathbf x}$
%the helical basis 
where $\alpha=(s_\alpha,\mathbf{k}_\alpha)$ is a multi-index carrying the chirality $s=\pm1$ and the wave number $\mathbf{k}\in \mathbb {Z}^{3}/L$. Accordingly, $v_\alpha(t)$ is the projection of $\bm{v}\leri{\bm{k}_\alpha,t}$ on $\bm{h}_{s_\alpha}$.
\begin{comment}
\begin{equation}
i\,\mathbf k\times\mathbf h_s(\mathbf k)=s\,k\,\mathbf h_s(\mathbf k),
\qquad
\mathbf h_s(\mathbf k)\cdot\mathbf h_{s'}^*(\mathbf k)=2\,\delta_{ss'}.
\end{equation}
The corresponding vorticity amplitudes are $\omega_s(\mathbf k,t)=s\,k\,v_s(\mathbf k,t)$. 
\end{comment}
The Euler equation becomes \cite{waleffe1992nature}
\begin{equation}
\partial_t v_\alpha
= \sum_{\beta,\gamma}
C^{\beta\gamma}_\alpha\,v^*_\beta\,v^*_\gamma ,
\label{eq:waleffe-alpha}
\end{equation}
where $C^{\beta\gamma}_\alpha= -\frac{1}{4}\big(s_\beta k_\beta - s_\gamma k_\gamma\big)\ g_{\alpha\beta \gamma}$, with $g_{\alpha \beta \gamma}= \mathbf{h}_{s_\alpha}^*(\mathbf k_\alpha)\cdot
\Big(\mathbf h^*_{s_\beta}(\mathbf k_\beta)\times 
      \mathbf h^*_{s_\gamma}(\mathbf k_\gamma)\Big)$  nonzero only if $\mathbf{k}_\alpha+\mathbf{k}_\beta+\mathbf{k}_\gamma=\mathbf{0}$.
In this formulation, the dynamics is organized into triadic interactions whose strength depends explicitly on the helicities of the participating modes, so that different polarization combinations produce qualitatively different nonlinear transfers. In this representation, the quadratic invariants are diagonal:
\begin{align}
E=\frac{1}{2}\sum_{\alpha}E_{\alpha}:=\frac{1}{2}\sum_{\alpha}v_{\alpha}v_{\alpha}^{*},\,\,\,\,\,\,H=\sum_{\alpha}S_{\alpha}E_{\alpha},
\end{align}
%$E=\frac{1}{2}\sum_{\alpha}v_{\alpha}v_{\alpha}^{*}$ and $H=\sum_{\alpha}S_{\alpha}v_{\alpha}v_{\alpha}^{*}$, 
where $S_{\alpha}:=s_{\alpha}k_\alpha$. Their conservation imposes the following restrictions on the interaction coefficients, 
\begin{align}\label{eq:C_con}
C_{\alpha}^{\beta\gamma}+C_{\beta}^{\alpha\gamma}+C_{\gamma}^{\alpha\beta}&=0,\\S_{\alpha}C_{\alpha}^{\beta\gamma}+S_{\beta}C_{\beta}^{\alpha\gamma}+S_{\gamma}C_{\gamma}^{\alpha\beta}&=0.
\end{align}
Since helicity is a Casimir of the algebra induced by the Poisson form, this might impose additional constraints on the nonlinear interactions. We do not pursue this question further here.

%\begin{equation}
%C^{\beta\gamma}_\alpha
%:= -\frac{1}{4}\big(s_\beta k_\beta - s_\gamma %k_\gamma\big)\,
%\mathbf h_{s_\alpha}^*(\mathbf k_\alpha)\cdot
%\Big(\mathbf h_{s_\beta}(\mathbf k_\beta)\times 
%      \mathbf h_{s_\gamma}(\mathbf k_\gamma)\Big).
%\label{eq:C-alpha}
%\end{equation}
%The skew symmetric operator takes the simple form $(\mathcal{J}\leri{\mathbf{a}})_\alpha=s_\alpha k_\alpha \sum_{\beta \gamma}g^{\beta \gamma}_\alpha\omega_\alpha a_\gamma$, where $\omega_\alpha=s_\alpha k_\alpha v_\alpha$. $(\mathcal{J}\leri{\mathbf{a}})_\alpha=\frac{s_\alpha k_\alpha}{2} \sum_{\beta, \gamma} s_\beta k_\beta a_{s_\beta}^* \omega_{s_\gamma}^* g_\alpha^{\beta \gamma}$.

\begin{comment}

The Lie–Poisson bracket $\{F,G\}_\mathcal{J}=\sum_{\alpha,\beta}
\frac{\delta F}{\delta \omega_\alpha}\,
\mathcal{J}_{\alpha\beta}[\omega]\,
\frac{\delta G}{\delta \omega_\beta},
$ takes a simple form in the helical basis 
\begin{equation}
\mathcal{J}_{\alpha\beta}[\omega]
=\sum_{\gamma}
C^{\alpha\beta}_\gamma\,
\frac{(s_\alpha k_\alpha)(s_\beta k_\beta)}{s_\gamma k_\gamma}\,
\omega_\gamma.
\label{eq:J-C}
\end{equation}
%\qquad J_{\alpha\beta}=-J_{\beta\alpha}
Here $\omega_\alpha=S_\alpha v_\alpha$, with $S_\alpha:=s_\alpha k_\alpha$.
Substituting $F(\omega)=\omega_\alpha$ and $G(\omega)=E[\omega]$ into this bracket reproduces Waleffe’s triad form \eqref{eq:waleffe-alpha}.
\end{comment}

\textbf{\textit{Dispersive waves and the kinetic equation}}
To introduce rotation- or chirality-induced dispersive wave dynamics, we add a linear %dispersive operator that breaks time parity 
dispersive operator of the form: $\nabla \cross iL(\frac{1}{i}\del\cross)\bf {v}$ to the %\ac{RHS} 
right-hand side of the Euler equation~\eqref{eq:euler}. This operator manifests non-Hermitian behavior due to its antisymmetric structure. However, since the operator is diagonal in terms of the curl eigenfunctions and $\omega_\alpha:=s_\alpha L(\bf{k})$ is real for $\mathbf{k}\in \mathbb{R}^{3}$, the conservation of energy and helicity persists. This is the case, for example, for a uniformly rotating fluid with 
\begin{equation}
    iL= -2\mathbf{\Omega}\times \mathbf{v},
\end{equation}
where the rotation axis is taken as $\mathbf{\Omega}=\Omega \hat{z}$. The dispersion relation is then
\begin{equation}
    \omega^\Omega_\alpha=\pm\Omega k_z/k.
\end{equation}
Another example arises in fluids with odd viscosity,
\begin{equation}\label{eq:oddvis}
    iL= \boldsymbol{\nu}^{\text{odd}}\times \Delta\mathbf{v},
\end{equation}
which can be viewed as a differential form of rotation \cite{de2024pattern}. In the simple case $\boldsymbol{\nu}^{\text{odd}}=\nu^{\text{odd}}\bm{\hat{z}}$, and 
\begin{equation} \label{eq:dispersion}
\omega_{\alpha}=\omega_{\left(s,\mathbf{k}\right)}=s_\alpha\nu^{\text{odd}}k_{z}k.
\end{equation}
The dispersive operator generates plane waves, which are solutions of the linear part of the Euler equation: $v_\alpha(t)=v_\alpha(t=0)e^{i\omega_\alpha t}$, where $v_\alpha(t=0)$ is the initial condition. Filtering the linear motion $v_\alpha\rightarrow e^{-i \omega_\alpha t}v_\alpha$, the Euler equation becomes 
\begin{align} \label{eq:eulerwave}
\partial_{t}v_{\alpha}=\sum_{\beta,\gamma}C_{\alpha}^{\beta\gamma}\,v_{\beta}^{*}\,v_{\gamma}^{*}e^{i\left(\omega_{\alpha}+\omega_{\beta}+\omega_{\gamma}\right)t}.
\end{align}
The oscillatory phase multiplying the nonlinear interaction introduces a separation of time scales in the evolution of the nonlinear solution for \textit{small initial data}. This separation allows one to derive a kinetic equation describing the slow evolution of the averaged wave component of the energy density $e_\alpha=\mean{E_\alpha}$. The overbar denotes averaging over an initial Gaussian statistical ensemble,
\begin{equation}\label{eq:cor}
\mean{v_\alpha^*(0)v_\beta(0)} = \delta_{\alpha\beta} e_\alpha(0),
\end{equation}
and its evolution is governed by the kinetic equation
\begin{align}
      \label{eq:kin sim}
  \dot \ea&=St_\alpha (e_\alpha)\\ \nonumber
&=\sum_{\beta,\gamma}\int_{\omega_{\alpha\beta\gamma}}C_{\alpha}^{\beta\gamma}\left(C_{\alpha}^{\beta\gamma}e_{\beta}e_{\gamma}+C_{\beta}^{\alpha\gamma}e_{\alpha}e_{\gamma}+C_{\gamma}^{\alpha\beta}e_{\alpha}e_{\mathbf{\gamma}}\right),
\end{align}
that was derived by \cite{de2025non}.
\begin{comment}
\begin{align}
      \label{eq:kin sim}
  \dot \ea&=St_\alpha (e_\alpha)\\ \nonumber
  &={\pi}\!\!
\int\limits_{\omega_{\alpha\beta\gamma}}\!\!\oma\,\Gamma_{\alpha\beta\gamma}^2 (\oma\eb\ec + \omb\ea\ec +
    \omg\ea\eb).%\delta(\omega_{\alpha \beta\gamma })
\end{align}
\end{comment}
To derive the kinetic equation, which provides the first nontrivial closure for the evolution of the wave component of the energy, one takes the joint kinetic limits of large domain and long nonlinear times, $L\rightarrow\infty$ and $t\omega_\alpha \rightarrow\infty$. In this limit, the discrete sum over the lattice in Eq.~\eqref{eq:waleffe-alpha}  is replaced by an integral over the resonant manifold:
\begin{equation}\label{eq:resonant manifold}
    \int\limits_{\omega_{\alpha\beta\gamma}} \!\!\!\!\!= \!\int \dd\beta \dd\gamma \,\delta(\omega_\alpha+\omega_\beta+\omega_\gamma)\delta\left(\mathbf{k}_{\alpha}+\mathbf{k}_{\beta}+\mathbf{k}_{\gamma}\right),
    \end{equation}
where $\int\!\! \dd\alpha=\sum_{s=\pm 1} \int\!\!\dd\mathbf{k}_\alpha$. \\
Near small frequencies, the kinetic equation must be interpreted carefully. In the discrete sum on the lattice, i.e., Eq.~\eqref{eq:eulerwave}, slow modes are well separated from waves with non-zero frequency. However, as $L\rightarrow \infty$, the collision integral~\eqref{eq:kin sim} includes integration arbitrarily close to slow modes. While slow modes cannot be created by resonant interactions, as $\omega_\alpha\rightarrow0$ the kinetic equation needs to include off-diagonal correlators apart from \eqref{eq:cor}, such as $\overline{v_\alpha^2}$, which would yield a very complicated description. Such correlators oscillate with frequency $\sim 2\omega_\alpha$ and over long times can be neglected as long as the frequency is bounded away from zero $|\omega_\alpha|\!>\!\epsilon\!>\!0$ \cite{shavit2024sign}. From this perspective, anisotropic systems whose dispersion relation vanishes along a curve rather than at a point pose a particular challenge for the kinetic description.

%\addMS{{Add comment on the relation to internal gravity waves, and we apply here a similar approach \cite{shavit2024sign}}}
On the resonant manifold, conservation of helicity can be used to simplify the kinetic equation as follows. Consider the vector of interaction coefficients for a triad $(\alpha,\beta,\gamma)$,
\begin{equation}
\left(C_{\alpha}^{\beta\gamma},C_{\beta}^{\alpha\gamma},C_{\gamma}^{\alpha\beta}\right)
\end{equation}
as a vector in $\mathbb{R}^{3}$. Conservation of energy and helicity implies that this vector is orthogonal to both
\begin{align}
(1,1,1), \qquad (S_\alpha,S_\beta,S_\gamma).
\end{align}
Except for degenerate cases, the three vectors
\begin{align}
(1,1,1), \qquad (S_\alpha,S_\beta,S_\gamma), \qquad , \left(k_{\alpha}^{z},k_{\beta}^{z},k_{\gamma}^{z}\right)
\end{align}
span $\mathbb{R}^3$. On the resonant manifold $\left(k_{\alpha}^{z},k_{\beta}^{z},k_{\gamma}^{z}\right)$ is also orthogonal to both $(1,1,1)$ and $(S_\alpha,S_\beta,S_\gamma)$. Therefore, the interaction coefficients must be parallel to $\left(k_{\alpha}^{z},k_{\beta}^{z},k_{\gamma}^{z}\right)$ and hence can be written in terms of a symmetric coupling $\Gamma_{\alpha\beta\gamma}$ as
\begin{align}
\left(C_{\alpha}^{\beta\gamma},C_{\beta}^{\alpha\gamma},C_{\gamma}^{\alpha\beta}\right)&=\Gamma_{\alpha\beta\gamma}\left(k_{\alpha}^{z},k_{\beta}^{z},k_{\gamma}^{z}\right)\,.
\end{align}
This reduces the kinetic equation to the simpler form
\begin{align}\label{eq:kinsimple}
\partial_{t}e_{\alpha}&= \sum_{\beta,\gamma}\int_{\omega_{\alpha\beta\gamma}}k_{\alpha}^{z}\Gamma_{\alpha\beta\gamma}^{2}\left(k_{\alpha}^{z}e_{\beta}e_{\gamma}+k_{\beta}^{z}e_{\alpha}e_{\gamma}+k_{\gamma}^{z}e_{\alpha}e_{\mathbf{\beta}}\right)\,.
\end{align}
Degenerate cases include for example $S_\alpha=S_\beta=S_\gamma$, in which case the interaction coefficients vanish.
The collision integral $St_\alpha$, which corresponds to the \ac{RHS} of this equation, admits the equilibrium solution $e_{\alpha}=1$. We now turn to finding solutions that carry a nonzero flux. \\

\textbf{\textit{Boundary conditions and steady solutions of the kinetic equation.}} The frequency %\eqref{eq:omega} 
and the geometric coupling %\eqref{eq:gamma} 
are symmetric under the dilation transformation: $\bm{k} \to \lambda \bm{k}$, for $\lambda>0$, where the homogeneity degree $x$ is determined by the scaling law $F(\lambda \bm{k}) = \lambda^x F(\bm{k}) $. That is, 
\begin{align}
\omega_{\left(s,\lambda\mathbf{k}\right)}&=\lambda^2\omega_{\left(s,\mathbf{k}\right)}, \\
\Gamma\left(\lambda\mathbf{k}_{\alpha},\lambda\mathbf{k}_{\beta},\lambda\mathbf{k}_{\gamma}\right)&=\Gamma\left(\mathbf{k}_{\alpha},\mathbf{k}_{\beta},\mathbf{k}_{\gamma}\right).
%V_{\lambda\mathbf{k}_{\alpha}}^{\lambda\mathbf{k}_{\beta}\lambda\mathbf{k}_{\gamma}}&=\lambda, V_{\mathbf{k}_{\alpha}}^{\mathbf{k}_{\beta}\mathbf{k}_{\gamma}}.
\end{align} %In particular, they are homogeneous functions of the wavenumbers  of
The wavenumbers themselves are homogeneous of degree one, and therefore the delta functions that define the resonant manifold are also homogeneous. Assuming the steady spectrum $e_\alpha$ shares this dilation symmetry leads to a separable homogeneous form $e_{\alpha}=C_0 k^{-w} e_{\alpha}^{\Omega}\left(\theta_{k},\phi_k\right)$,  
%\begin{align} \label{eq:turbspec0}
%e_{\alpha}=C_0 K^{-w} e_{\alpha}^{\Omega}\left(\theta_{k},\phi_k\right), 
%\end{align} 
where $C_0>0$ is the Kolmogorov constant. Because the kinetic equation has azimuthal symmetry about the vertical axis, the angular dependence reduces to a function of the polar angle only. Motivated by the fact that the dispersion relation depends on the angle only through $\omega_\alpha/ \leri{\nu_\text{odd} s_\alpha k^2} = \cos\theta_k$, %: an SO(2) symmetry about the $\hat{z}$ axis, 
while the interaction coefficient is isotropic, we represent the spectrum as
\begin{align} \label{eq:turbspec}
e_{\alpha}=C_0 k^{-w} |{\cos \theta_k}| ^{-f_{w,\alpha}(\theta_k)}\,.
\end{align}
Substituting this homogeneous spectrum into the collision integral turns it into a homogeneous function that separates into radial and angular parts. The collision integral takes the form
 \begin{equation}
  St_\alpha\equiv k^{2w_0-2w-d}C_0^2St_{\Omega,\alpha}\left(w,f_w, \theta_k \right),
 \end{equation}
where $d=3$ is the dimension and $2w_0-2w-d$ is the total homogeneity degree of the elements entering the kinetic equation, written in this form for later use. In our case $w_0=d=3$. The explicit form of the angular collision integral, $St_\Omega$, is obtained by parametrizing the resonant manifold and is presented in the supplemental material.

The separability of the collision integral suggests the existence of a one-parameter family of scale-invariant spectra parametrized by the exponent $w$. To select the physically relevant solution, we impose boundary conditions through the energy flux.
Since the Energy is an integral of motion, the kinetic Eq.~\eqref{eq:kinsimple} %$\dot{e}\left(\mathbf{k}\right)=St\left(\mathbf{k},e\right)$
can be written formally as a continuity equation 
\begin{align}
\dot{e_\alpha}\left(t\right)+\text{div}\Pi_\alpha\left(t\right) & =0,
\end{align}
where $\Pi_\alpha\left(t\right)$ %$\Pi_\alpha\left(t\right)=\left(\Pi^r_\alpha(t),\Pi^{\theta}_\alpha(t)\right)$
is a three-dimensional flux in k-space. Its divergence is related to the collision integral through
\begin{align}\label{eq:divSt}
\text{div}\Pi_\alpha(t) =-St_\alpha (e_\alpha)\,.
\end{align}
\addMS{Accordingly, stationary solutions correspond to divergence-free fluxes. The flux field is not uniquely determined by its divergence: if $A_\alpha$ is any vector potential, then $\Pi_\alpha\rightarrow\Pi_\alpha+\nabla_k\times A_\alpha$,
leaves \eqref{eq:divSt} unchanged. Thus, infinitely many flux fields correspond to the same stationary state. Conversely, specifying the flux uniquely determines the stationary solution. In stationary turbulence sustained between well-separated forcing and dissipation scales, a natural gauge choice is a purely radial flux, $\Pi_\alpha=\Pi_\alpha^{r}\hat{r}$, where the radial density $\Pi_\alpha^{r}$ may still depend on the polar angle, consistent with anisotropy. This choice is further motivated by the absence of resonant interactions among triads satisfying  $k_\alpha =k_\beta =k_\gamma$, with differing propagation directions only. Consequently, there is no purely angular resonant redistribution of energy at fixed wavenumber magnitude.}
With this choice, the divergence relation becomes
\begin{align}
\text{div}\Pi_\alpha & =k^{-d+1}\partial_k (k^{d-1}\Pi^r_\alpha) \equiv -St_\alpha(e_\alpha).
\end{align}
Substituting the homogeneous spectrum~\eqref{eq:turbspec} gives
\begin{align}\label{eq:wsolution}
k^{-d+1}\partial_{k}(k^{d-1}\Pi^r_{\alpha}) & =-k^{2w_0-2w-d}C_0^2St_{\Omega}\left(w,f_w, \theta_k \right)\,,
\end{align}
such that in a steady state, $St_{\Omega}\left(w_0,f_{w_0}, \theta_k \right)$ must vanish. Multiplying by $k$ and integrating $\int_k^\infty dk' $ assuming $w<w_0$ yields the radial flux  
\begin{align}\label{eq:flux}
k^{d-1}\Pi^r_{\alpha} & =-\frac{k^{2\left(w_{0}-w\right)}}{2\left(w_{0}-w\right)} C_0^2St_{\Omega}\left(w,f_w, \theta_k \right).
\end{align}
Integrating $\int_0 ^k dk'$ would yield the same expression for $w>w_0$. From this expression, it becomes clear that the pair $(w_0,f_{w_0})$ corresponds to a special solution: the unique scale-invariant spectrum that gives a non-zero\, scale-independent radial flux with zero divergence. This fixes the isotropic scaling exponent of the turbulent spectrum to
\begin{equation}\label{eq:spectrum}
    e_{\alpha}=C_0 k^{-3} \abs{\cos \theta_k} ^{-f_{w_0,\alpha}(\theta_k)}.
\end{equation}
As $w$ tends to $w_0$, due to the diverging denominator in Eq.~\eqref{eq:flux}, the radial flux is given by: %\addSAS{(added $C_0^2$ below:)}
\begin{align} \label{eq:fluxang}
\Pi^r_\alpha(w_0,f_{w_0}) =\frac{C_0^2}{2k^{d-1}}\frac{d}{dw}St_{\Omega,\alpha}\left(w,f_{w_0},\theta_k \right)\mid_{w=w_{0}}\,.
\end{align}
The derivative with respect to $w$ in the definition of the flux is taken after the angular part $f_w=f_{w_0}$ has been fixed in the angular collision integral. Isotropic wave turbulence~\cite{ZLF} is a particular case of Eq.~\eqref{eq:fluxang} where $St_{\Omega}\left(w,f_{w_0},\theta_k 
\right)=St_{\Omega}\left(w \right)$ is independent of the angle. In non-isotropic systems, it is natural to expect the flux $\Pi_\alpha^r$ to depend on the angle.  The average over a sphere gives the constant non-zero total radial flux
\begin{align}
2\pi\int_{0}^{\pi}k^{d-1}\Pi^r_\alpha \left(\theta\right)\sin\theta d\theta =\Pi_{0}.
\end{align}

The scaling obtained above arises in two distinct helicity configurations. In the first, helicity is equally distributed between left- and right-handed waves, so that the mean helicity vanishes: $e_{+,\mathbf{k}}=e_{-,\mathbf{k}}$ and $\overline{H}=0$. In the second, helicity is concentrated on a single branch, corresponding to a sign-definite helicity state, e.g $e_{-,\mathbf{k}}=0$, so that $\overline{H}>0$, \addMS{and the dynamics is restricted to the positive branch only}. 
In the latter case, the kinetic equation \addMS{restricted to the positive branch} admits an additional scaling solution corresponding to a helicity cascade. Since the helicity density at each wavenumber scales as $ke_{(+,k)}$, the flux relation is modified accordingly
\begin{align}\label{eq:flux_helicity}
k^{d-1}\Pi^{r,H}_{\alpha} & =-\frac{k^{2\left(w_{0}-w\right)+1}}{2\left(w_{0}-w\right)+1} C_0^2St_{\Omega}\left(w_H,f_{w_H}, \theta_k \right)\,,
\end{align}
and leads to a helicity-cascade scaling exponent $w_h=3.5$. The corresponding spectrum is
\begin{equation}\label{eq:helspectrum}
    e^H_{\alpha}=C_0 k^{-3.5} \abs{\cos \theta_k} ^{-f_{w_h,\alpha}(\theta_k)}.
\end{equation}
This spectrum is not an exact steady solution of the kinetic equation but rather an asymptotic regime that may arise near sources of chirality-definite waves. \addMS{To illustrate this, decompose the helicity into positive and negative components as $H =H_{+}+H_{-}$ with $H_+\geq0$ and $H_-\leq0$ and consider sign-definite initial data consisting entirely of positive-helicity waves, i.e $v_{(-\mathbf{k})}(0)=0$ for all $\mathbf{k}$.  Then $H=H_+(0)>0$ initially and in fact at all times by the exact conservation of $H$. Moreover, it follows from \eqref{eq:waleffe-alpha} that in this case the initial time derivatives $\dot{H}_{+}=\dot{H}_{-}=0$, indicating short-time persistence of the sign-definite helicity state. Although nonlinear interactions eventually generate waves of opposite helicity, conservation of total helicity constrains this process: any growth of negative-helicity modes must be accompanied by a compensating increase in positive-helicity modes. Consequently, one expects a persistent imbalance $H_+\gg|H_-|$ for all $t\geq0$, so that the dynamics remains dominated by the positive-helicity branch.}

Whether such a state can be realized physically or not, the existence of a scaling solution supported on the positive branch has important consequences for the distribution of energy flux among different helicity triads, to which we now turn.

\textbf{\textit{Helicity and flux directions.}} 
The density of helicity is proportional to the energy density through the isotropic and monotonic function $k$. As a consequence, the power-law exponent associated with the energy cascade is smaller than the exponent corresponding to the helicity cascade solution of the kinetic equation restricted to a single (positive) helicity branch, i.e., $w_0<w_h$.

Suppose that the angular part of the collision integral has the same kernel for both cascades, namely $f_{w_0}=f_3=f_{3.5}=f_{w_h}$. In this case, sign-definite helicity interactions would transfer energy toward smaller absolute wavenumbers (an inverse energy cascade), while sign-indefinite helicity interactions would transfer energy toward larger absolute wavenumbers (a direct cascade). This conclusion follows from a generalization of the flux analysis for isotropic scale-invariant solutions described in \cite{nazarenko2011wave}. Consider the family of spectra given by Eq.~\eqref{eq:turbspec}, and fix the angular dependence to be $f_w=f_{w_0}$, while allowing the homogeneous scaling $k^{-w}$ to vary. Within this restricted family, our analysis shows that only one spectrum corresponds to a steady state, namely $w=w_0=3$. All other spectra are non-stationary, but a flux can still be defined for them—whenever it is finite—by continuity from the expression derived earlier, \eqref{eq:flux}, which we denote by $\Pi_\alpha(w)$. 

In the limit $w\rightarrow \infty$, where most of the energy is concentrated at very small wavenumbers, the flux is expected to be positive. Since there are no additional steady solutions $w\neq w_0$ associated with other positive-definite invariants, the flux does not change sign near the energy-cascade solution and therefore remains positive at $w=w_0$. This corresponds to a forward (direct) energy cascade.

Now consider the kinetic equation restricted to interactions involving only positive-helicity waves. In this case, in addition to the energy-cascade solution $w_0$, the kinetic equation admits another scale-invariant solution corresponding to a helicity cascade at $w_h=3.5$. Because this spectrum is an exact solution on the positive branch, the energy flux must vanish there, as follows directly from the flux expression derived earlier.

Since $w_h>w_0$ and there are no additional zeros of the flux, the function $\Pi_\alpha(w)$ crosses zero once—starting from positive values as $w\rightarrow\infty$ and vanishing at $w=w_h$. It therefore remains negative at $w=w_0$. This implies that, within the helicity-definite dynamics, the energy flux at the energy-cascade scaling is directed toward larger spatial scales.

In the next section, we introduce an approximation motivated by the accumulation of energy near slow modes. Under this approximation, we show that the same behavior arises in the full odd-wave system. Thus, although helicity is globally sign-indefinite, the interactions involving helicity-definite triads constrain the cascade dynamics. In particular, when both positive- and negative-helicity waves are present, helicity-definite interactions drive a portion of the energy flux toward large scales.

\textbf{\textit{Slow modes and the angular part of the solution.}}
While the self-similar, scale-invariant part of the solution is largely insensitive to the precise domain of validity of the kinetic equation, the angular structure of the spectrum is not. This sensitivity arises from the limit $\theta_k\rightarrow\pi/2$, which corresponds to $\omega_\alpha\rightarrow0$. In this limit, the derivation of the kinetic equation breaks down, because the separation of linear and nonlinear time scales underlying weak turbulence theory no longer holds. Since waves with vanishing frequency correspond to slow modes, and since energy is expected to accumulate near these modes, the solution should exhibit a singular behavior along the curve of zero frequency, $k_\alpha^z=0$ or $\theta_k=\pi/2$.

We now determine $f_{w_0}$, the exponent of the angular part of the spectrum, and focus on solutions with a constant exponent $f$. Because the kinetic equation is independent of the azimuthal angles, we first integrate over them. This produces the averaged delta function
\begin{align}
    \int \dd\phi_{\beta}\int \dd\phi_{\gamma}\delta\left(\mathbf{k}_{\alpha}+\mathbf{k}_{\beta}+\mathbf{k}_{\gamma}\right)&=\frac{\delta\left(k_\alpha^{z}+k_\beta^{z}+k_\gamma^{z}\right)}{2 A_{2D}},
\end{align}
where $\phi$ are the cylindrical angles and $A_{2D} \equiv \frac{1}{4} \sqrt{4 \rho_\beta^2 \rho_\gamma^2-\leri{\rho_\beta^2 +\rho_\gamma^2 - \rho_\alpha^2}^2}$, 
is the area of the two-dimensional projection of the triangle formed by the resonant triad $\mathbf{k}_{\alpha}+\mathbf{k}_{\beta}+\mathbf{k}_{\gamma}=0$. Here $\rho_\alpha=\sqrt{\leri{k_\alpha^x}^2+\leri{k_\alpha^y}^2}$ denotes the projection of the absolute wave number $k$ on the plane $k_z=0$. The remaining integration of the collision integral is restricted to the domain $A_{2D}\geq 0$. Details of the calculation are given in the appendix. 

Because the spectrum is expected to be strongly concentrated near slow modes $k_\alpha^z=0$, we approximate the averaged two-dimensional projected area of the resonant triangle by its three-dimensional counterpart. More precisely, we assume that the average of $A_{2D}$ with respect to the measure induced by the energy density is well approximated by the corresponding average of the three-dimensional triangle area $A_{3D}$. \addSAS{We define the averaged N-dimensional triangle area by
\begin{align}
    \obs{A_{ND}}_{\beta \gamma} &= \int \dd V_\beta \dd V_\gamma A_{ND} e_\beta e_\gamma \delta \leri{\mathbf{k}_\alpha+\mathbf{k}_\beta+\mathbf{k}_\gamma } \,,
\end{align}
with $A_{3D} \equiv \frac{1}{4} \sqrt{4 k_\beta^2 k_\gamma^2-\leri{k_\beta^2 +k_\gamma^2 - k_\alpha^2}^2}$ and $A_{2D}$ the area of its projection onto the $k_z=0$ plane. Throughout the derivation of the angular component of the energy spectrum we approximate 
\begin{equation}
    \obs{A_{2D}}_{\beta \gamma} \sim \obs{A_{3D}}_{\beta \gamma},
\end{equation}
an approximation that becomes asymptotically exact in the strongly anisotropic slow-mode limit. Its validity is discussed in a dedicated section below and quantified numerically in the appendix. } Under this approximation, the collision integral becomes 
\begin{align}    
\partial_{t}e_{\alpha}&=\sum_{\beta,\gamma}\int \dd\!\cos\theta_{\beta}\dd\!\cos\theta_{\gamma}k_{\beta}^{2}dk_{\beta}k_{\gamma}^{2}dk_{\gamma}k_\alpha^z 
    % k_\beta \dd k_\beta \dd k_\beta^z k_\gamma \dd k_\gamma \dd k_\gamma^z k_\alpha^z 
    \nonumber \\
    &\frac{\Gamma_{\alpha\beta\gamma}^{2}}{2A_{3D}}\left(k_\alpha^z e_{\beta}e_{\gamma}+k_\beta^z e_{\alpha}e_{\gamma}+k_\gamma^z e_{\alpha}e_{\mathbf{\beta}}\right)\nonumber \\
    &\delta\left(\omega_{\alpha}+\omega_{\beta}+\omega_{\gamma}\right)\delta\left(k_\alpha^z+k_\beta^z+k_\gamma^z\right).
\end{align}

\begin{comment}
    \begin{align}    
\partial_{t}e_{\alpha}&=\sum_{\beta,\gamma}\int_{\Theta\left(A_{3D}\right)} d\cos\theta_{\beta}d\cos\theta_{\gamma}k_{\beta}^{2}dk_{\beta}k_{\gamma}^{2}dk_{\gamma}k_\alpha^z 
    % k_\beta \dd k_\beta \dd k_\beta^z k_\gamma \dd k_\gamma \dd k_\gamma^z k_\alpha^z 
    \nonumber \\
    &\frac{\Gamma_{\alpha\beta\gamma}^{2}}{2A_{3D}}\left(k_\alpha^z e_{\beta}e_{\gamma}+k_\beta^z e_{\alpha}e_{\gamma}+k_\gamma^z e_{\alpha}e_{\mathbf{\beta}}\right)\nonumber \\
    &\delta\left(\omega_{\alpha}+\omega_{\beta}+\omega_{\gamma}\right)\delta\left(k_\alpha^z+k_\beta^z+k_\gamma^z\right),
\end{align}

where $\Theta\left(A_{3D}\right)$ is the Heaviside function that limits the integration on the set of wavenumbers where the area of the triangle is non-negative. 
\end{comment}
Substituting the spectral ansatz introduced earlier \eqref{eq:turbspec}, the kinetic equation is separable in the variables $k$ and $\cos\theta$.
The leading-order singular part of the reduced angular collision integral is locally homogeneous in $\cos\theta$ as $\cos\theta \rightarrow 0$. The two spectral factors contribute the singular degree $-2f_w$, while the remaining contributions are scale invariant to leading order.
Treating the angular spectrum as a generalized zero of the angular collision integral in the principal-value sense around $\theta_k=\pi/2$, yields $2f_w=-1$. This determines the angular part of the energy spectrum
\begin{equation}\label{eq:spectrum_final}
    e_{\alpha}=C_0 k^{-3} |\cos \theta_k| ^{-1/2}.
\end{equation}
This spectrum is illustrated in Fig.~\ref{fig:2dspec}, in terms of the wavenumber cylindrical coordinates: $e_\alpha = C_0 \leri{\rho_\alpha^2 + {k_\alpha^z}^2}^{-5/4} \abs{k_\alpha^z}^{-1/2} $. For the helicity cascade on the positive branch, the angular exponent remains the same, $f_{w,h}=f_{w_0}$, giving 
\begin{equation}\label{eq:hspectrum_final}
    e^{H}_{\alpha}=C_0 k^{-3.5} |\cos \theta_k| ^{-1/2}.
\end{equation}
Under the approximation described above, sign-definite helicity interactions therefore drive a portion of the energy flux toward large scales, consistent with the mechanism discussed in the previous section.

For comparison, ~\cite{de2025non}, de Wit et al. derived the turbulent energy spectrum assuming the strongly anisotropic limit where $k$ is replaced by its planar projection $\rho$ through the kinetic equation. This renders the kinetic equation separable in $\rho$ and $k_z$ and leads to the scaling $e_\alpha\propto \rho^{-2.5}|k_z|^{-0.5}$. In spherical coordinates this corresponds to $e_\alpha\propto k^{-3}|\cos \theta_k|^{-0.5}|\sin \theta_k|^{-2.5}$, which contains an unphysical divergence along $\theta=0$.

The spectrum \eqref{eq:spectrum_final} avoids this pathology, and separates into two parts: an exact isotropic scaling $k^{-3}$, which follows directly from the structure of the kinetic equation without additional assumptions, and an angular part $|\cos\theta_k|^{-1/2}$, obtained under the mild approximation introduced above. This yields a spectrum that remains regular away from the slow-mode curve while capturing the expected energy accumulation near $\theta=\pi/2$.

\textit{\textbf{Implications to inertial waves}}.
We now consider the case of inertial waves in a rapidly rotating fluid. 
Unlike the case of odd waves, the vector of frequencies $(\omega_\alpha,\omega_\beta,\omega_\gamma)$ on the resonant manifold is orthogonal both to to $(1,1,1)$ and the helicity vector $(S_\alpha,S_\beta,S_\gamma)$. As a consequence, the interaction coefficients simplify to
\begin{align}
\left(C_{\alpha}^{\beta\gamma},C_{\beta}^{\alpha\gamma},C_{\gamma}^{\alpha\beta}\right)&=\Gamma_{\alpha\beta\gamma}\left(\omega_\alpha,\omega_\beta,\omega_\gamma\right)\,,
\end{align}
And thus, in this case the kinetic equation takes the form
\begin{align}\label{eq:kinsimple2}
\partial_{t}e_{\alpha}&= \sum_{\beta,\gamma}\int_{\omega_{\alpha\beta\gamma}}\omega_{\alpha}\Gamma_{\alpha\beta\gamma}^{2}\left(\omega_{\alpha}e_{\beta}e_{\gamma}+\omega_{\beta}e_{\alpha}e_{\gamma}+\omega_{\gamma}e_{\alpha}e_{\mathbf{\beta}}\right)\,.
\end{align}
%Despite this structural difference in the interaction coefficients, 
The turbulent spectrum obtained under the approximation introduced earlier $\obs{A_{2D}}_{\beta \gamma} \sim \obs{A_{3D}}_{\beta \gamma}$ for the projected triangle area is given by
\begin{equation}\label{eq:spectrum_final} 
    e_{\alpha}=C_0 k^{-4} |\cos \theta_k| ^{-1/2}.
\end{equation}
The same weak singularity as for the case of odd waves $|\cos \theta_k| ^{-1/2}$ reflects the concentration of energy near the slow-mode manifold $k_z \approx 0$,  consistent with the strongly anisotropic behavior predicted in the classical theory of inertial-wave turbulence in rotating fluids \cite{galtier2003weak, Cambon1997, Cambon2004}.

The same reasoning also applies to the helicity cascade on the positive branch. In particular, the helicity cascade solution and the conclusion that sign-definite helicity interactions drive a portion of the energy flux toward large scales remain valid for inertial waves as well.

{\textbf{\textit{Validity of the projected-area approximation}}
\addSAS{ Using $\mathbf{k}_\alpha+\mathbf{k}_\beta+\mathbf{k}_\gamma=0$ both triangle areas are cross products of a single pair of legs,
\begin{align}
A_{3D}=\tfrac12\abs{\mathbf{k}_\alpha\times\mathbf{k}_\beta}\,, \qquad A_{2D}=\tfrac12\abs{\leri{\mathbf{k}_\alpha\times\mathbf{k}_\beta}_z}\,,
\end{align}
so that their ratio measures the tilt of the triad plane relative to $\hat z$. The fidelity of the approximation is set by
\begin{align} \label{eq:areaRatio}
    R(\theta_\alpha) \equiv \frac{\obs{A_{3D}}_{\beta \gamma}}{\obs{A_{2D}}_{\beta \gamma}},
\end{align}
which measures the average inclination of the triad plane relative to the $k_z=0$ plane. Since $A_{2D}$ is the projection of the full area, $R\geq 1$, with equality only when the triad lies in the horizontal
plane.
For the anisotropic spectrum
$
e_k\propto |\cos\theta_k|^{-1/2},
$
the weight $e_\beta e_\gamma$ strongly favors partner wavevectors close to the slow manifold. Consequently, as
$k_\alpha^z\rightarrow0$,
all three wavevectors become nearly coplanar with the horizontal plane, implying
\begin{align}
R(\theta_\alpha)
\xrightarrow[k_\alpha^z\to0]{}
1.
\end{align}
The projected-area approximation
$\langle A_{2D}\rangle_{\beta\gamma}\simeq
\langle A_{3D}\rangle_{\beta\gamma}$
therefore becomes asymptotically exact precisely in the region where the energy is concentrated, see Figure ~\ref{fig:triangleAreas}.
Importantly, this agreement is not a purely geometric property but a consequence of the anisotropic weighting. For an isotropic spectrum, for example
$e_k\propto k^{-3}$,
the partner wavevectors retain generic inclinations even as
$k_\alpha^z\rightarrow0$, yielding instead
\begin{align}
R(\theta_\alpha) = \frac{\pi}{2 \sin \theta_\alpha}\rightarrow _{k_z\rightarrow 0 }\frac{\pi}{2}\approx1.57, 
\end{align}
After averaging over the remaining leg, the integrated ratio
$\langle A_{3D}\rangle/\langle A_{2D}\rangle$
equals $2$ for an isotropic distribution, a distribution-independent value determined by the mean orientation of the triad plane. For the odd-wave spectrum, the corresponding ratio decreases to approximately $1.31$ (see the Appendix \cite{SM}), reflecting the strong concentration of the measure near the slow manifold.
The monotonic decrease of $R$ toward unity as the distribution becomes increasingly anisotropic demonstrates that the projected-area approximation is controlled in the regime relevant to the present theory, while its accuracy deteriorates continuously as the distribution approaches isotropy.}
%This is a genuine consequence of the anisotropy: for an isotropic distribution, e.g $e_{k}\propto k^{-3}$ or any other isotropic distribution, the partner legs retain generic inclinations even on the slow manifold, and the same limit yields the finite value
%\begin{align}
%R(\theta_\alpha) = \frac{\pi}{2 \sin \theta_\alpha}\rightarrow _{k_z\rightarrow 0 }\frac{\pi}{2}\approx1.57, 
%\end{align}}

%and for an isotropic distribution the normal \hat{\mathbf w} is uniform on the sphere, giving ⟨|\cosψ|⟩=\tfrac12 and the ratio =2.

\begin{figure}[t]  
\includegraphics[width=0.5\textwidth]{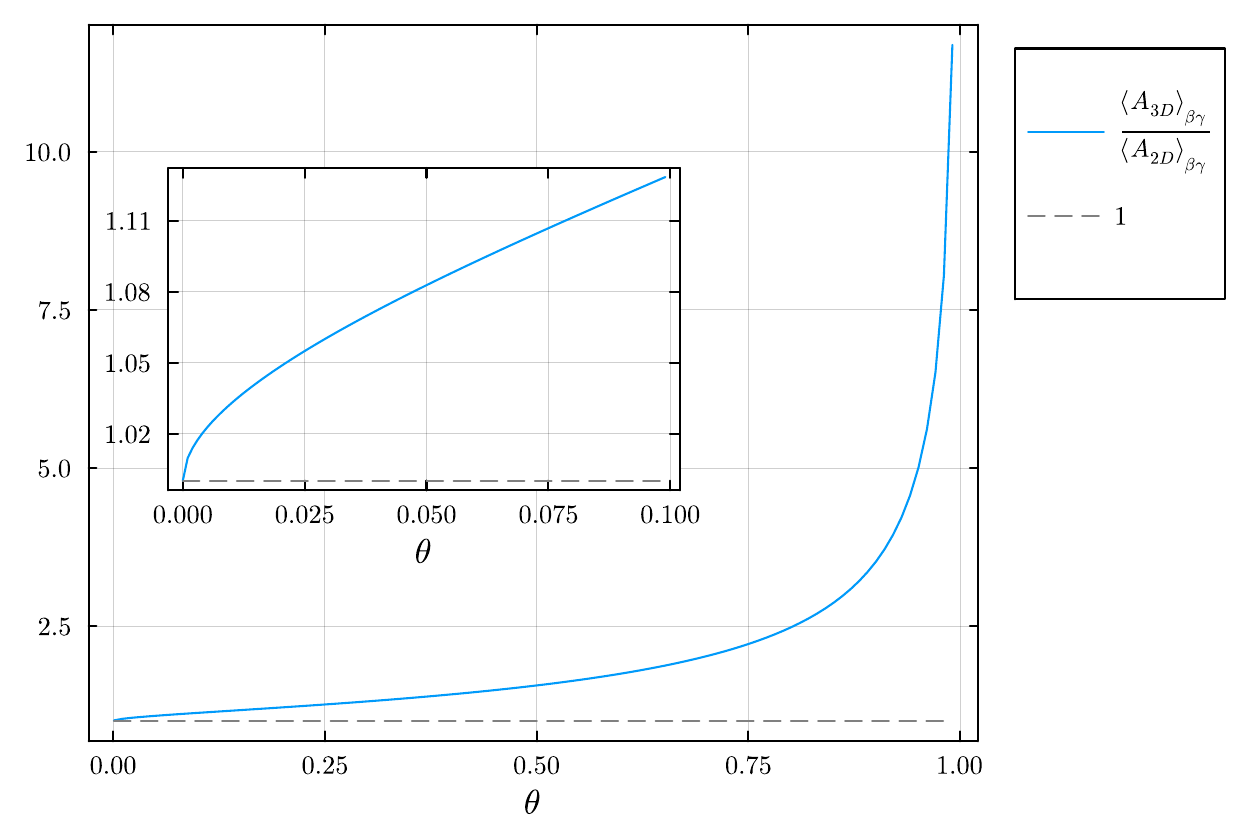}
\caption{The approximation $\obs{A_{3D}}_{\beta \gamma} \sim \obs{A_{2D}}_{\beta \gamma}$, as defined in Eq.~\eqref{eq:areaRatio}. We present the numerical ratio between the averaged area of a triangle in 3D and 2D (blue line), as a function of $\theta$. Inset: We focus on low values of $\theta$ which correspond to slow modes, and note that the ratio tends to unity.} \label{fig:triangleAreas} %\shahaf{power-law error?}. Numerical parameters: $hu=40,\ h=d=8$, and absolute tolerance given by 1e-5.}
\end{figure}

\textbf{\textit{Family of solutions in the very anisotropic limit and numerical evaluation of our findings}}
We now demonstrate numerically the results described above in the standard strongly anisotropic limit, where the magnitude of the wavenumber is approximated by its planar projection, $k\sim\rho$. 

The normalized angular collision integral $\cos\theta_kSt_\Omega$, is homogeneous of degree zero in the angular variable $\cos\theta$. We verify numerically that the pair $(w_0,f_{w_0})$ corresponds to a zero of this collision integral.

In addition to the turbulent solution, the numerical evaluation reveals a one-dimensional family of approximate solutions of the form  $e_\alpha=k^{-w}|\cos\theta_k|^{-f_w}$, with parameters approximately in the range $w\in{[2.8,3.2]}$ and $f_w\in[0.44,0.57]$. The turbulent solution lies within this family at the point $(w_0,f_{w_0})=(3,0.5)$. The set of zeros of the collision integral in the $(w,f)$ plane is presented in Figure~\ref{fig:findZero}. 

\begin{figure}[t] 
\includegraphics[width=0.5\textwidth]{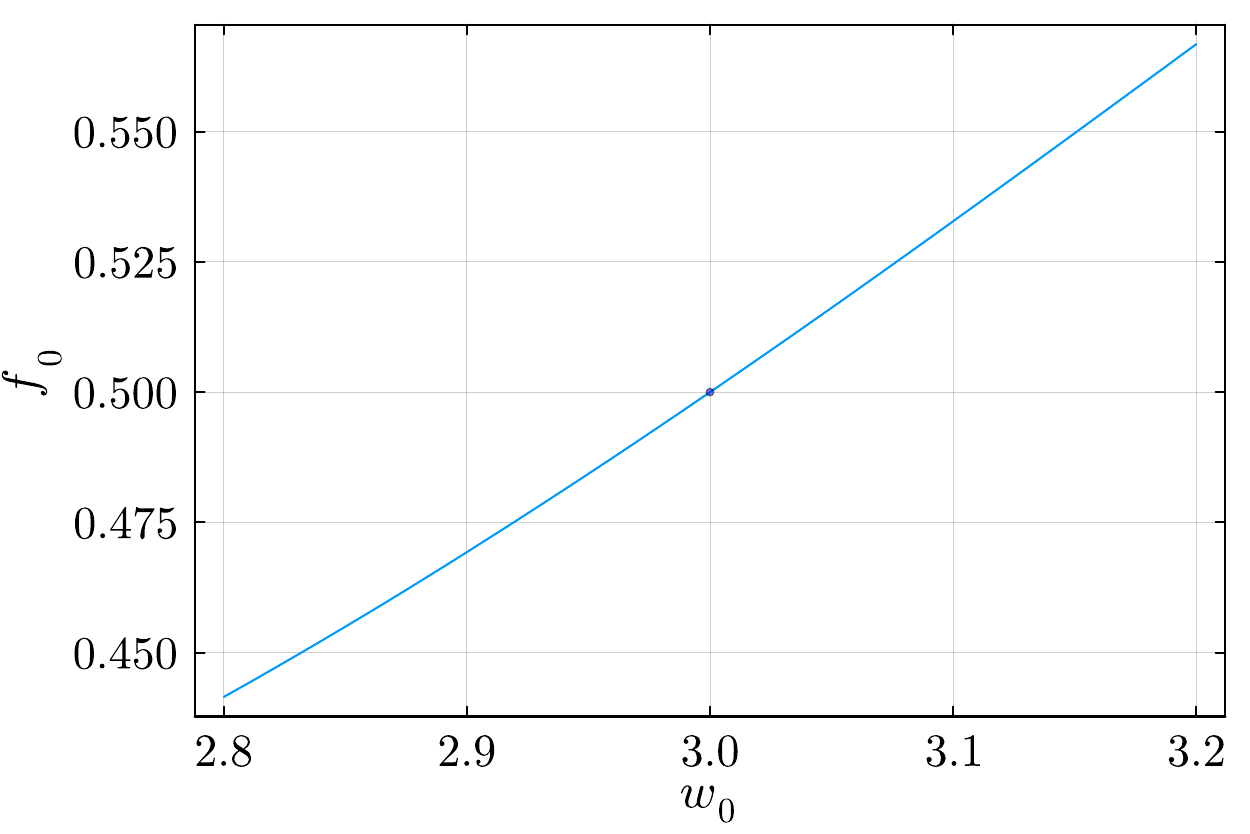}
\caption{ Set of zeros of the collision integral, found numerically~\cite{Roots.jl} in terms of the angular $f_0$ and radial $w_0$ power-laws. We highlight the energy cascade solution: $w_0=3,\ f_0=0.5$.}\label{fig:findZero}
\end{figure}

Next, we verify that sign-definite helicity interactions among waves belonging to the same helicity branch with $s_\alpha=s_\beta=s_\gamma$ transfer energy backward in wave number $k$.  That is, the contribution to the flux $\Pi_\alpha$ from interactions among the same branch is negative: $\left. {\Pi_\alpha}%
 \right|_{(+,+,+)}\!<\!0$ for all $\omega_k$. Normalized by the total flux, the contribution of triads $s_\alpha\neq s_\beta=s_\gamma$ is the smallest, and the contribution of the triads $s_\alpha=s_\beta\neq s_\gamma$ and $s_\alpha=s_\gamma\neq s_\beta$ is positive to a forward cascade, so that $\left. {\Pi_\alpha}%
 \right|_{(+,-,+)}\!=\!\left. {\Pi_\alpha}%
 \right|_{(+,+,-)}\!>0\!$. %We also have the following distribution of contributions to the angular dependent flux $\left. {\Pi_\alpha}% 
 %\right|_{(+,+,+)}/\left. {\Pi_\alpha}%
 %\right|_{(+,-,+)}=-\frac{1}{2}$. So the fraction of energy transported backwards is $1/3$ of the total energy, and is transported by the set of waves belonging to the same branch on which the pseudo-momentum is sign definite. 
 %, such as the Hawaiian ridge \cite{LSY03}.  
\addSAS{The total energy flux $\Pi_\alpha$ (green), and the fluxes restricted to the different triads: positive $\left. {\Pi_\alpha}%
 \right|_{(+,+,+)}$ (pink), negative $\left. {\Pi_\alpha}%
 \right|_{(+,-,-)}$ (brown), and mixed $\left. {\Pi_\alpha}%
 \right|_{(+,-,+)} + \left. {\Pi_\alpha}%
 \right|_{(+,+,-)}$ (blue) are plotted in Figure~\ref{fig:fluxes}.}

%This suggests the existence of an inverse energy cascade of odd (and inertial) waves when interactions among waves of different branches are suppressed, for example, in the vicinity of a source emitting helicity-definite waves.

 These results reveal a clear separation between the roles of helicity-definite and helicity-mixed interactions. Triads composed of waves belonging to the same helicity branch drive energy toward larger spatial scales, while interactions involving waves of opposite helicity branches sustain the forward cascade. Consequently, although helicity is globally sign-indefinite, the decomposition into helicity branches introduces sign-definite interaction channels that systematically transfer a portion of the energy toward large scales.

This mechanism implies that when interactions between helicity branches are weakened—for example, near a source that emits helicity-definite waves—the dynamics can favor an inverse energy transfer. More generally, the results demonstrate that helicity organizes the cascade structure at the level of resonant triads, producing systematic backscatter even in regimes where the net energy cascade remains direct.

\begin{figure}[t] 
\includegraphics[width=0.5\textwidth]{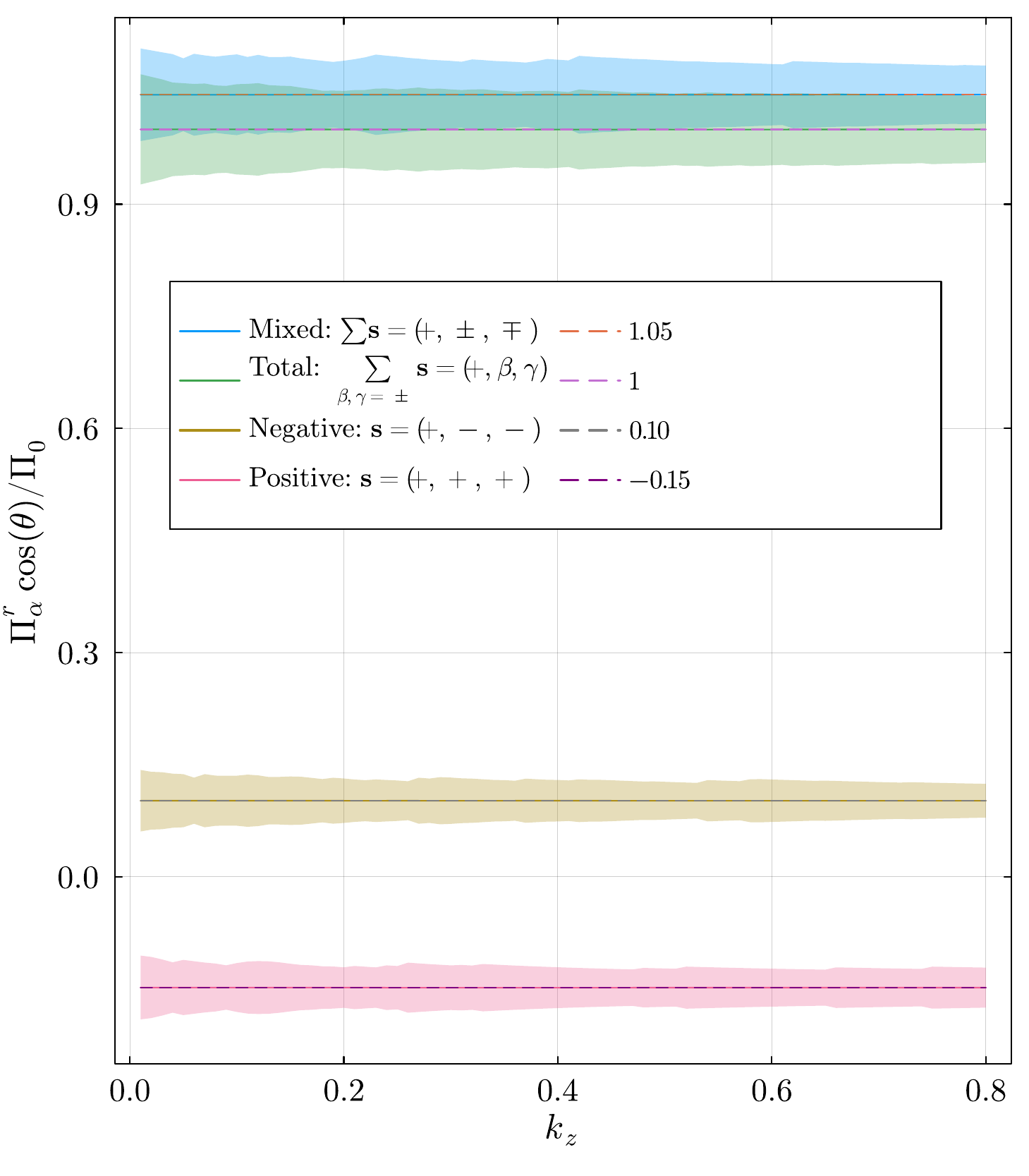}
%{PRL/energyFluxesVsKzPlot_Normalized_v2.pdf}
\caption{\addSAS{Radial energy fluxes of different branches (colors)}, as given by Eq.~\eqref{eq:fluxang} (See appendix for the explicit form), as a function of $k_\alpha^z$, evaluated at $k_\alpha=1$. The fluxes were numerically integrated~\cite {HCubature} and normalized to the total energy flux (green line). Error regions correspond to $\pm 100$ error estimation of the numerical integration.} \label{fig:fluxes}
\end{figure}

\textit{\textbf{Conclusion and perspectives.}}

We studied weak turbulence in incompressible flows with broken time-reversal symmetry. Our primary example is the case of odd-viscous (“odd”) waves, but all of our analytical arguments and conclusions extend directly to rotating Euler flow (inertial waves), with the corresponding substitutions in the dispersion relation and resonant geometry. In both systems, the wave component admits a kinetic description whose collision integral inherits two quadratic invariants—energy and helicity—but reorganizes them in a strongly anisotropic manner. This setting provides a natural framework for understanding how a sign-indefinite invariant can nevertheless impose constraints on turbulent cascades.

Our first main result is a flux-based selection principle for the steady, scale-invariant spectrum. By introducing a natural gauge fixing for the energy flux, choosing a purely radial flux in $\mathbf{k}$ space, we showed that among the dilation-invariant family $e_\alpha=C_0 k^{-w}e^\Omega(\theta_k)$ there is a unique exponent supporting a nonzero, divergence-free radial energy flux in the inertial range: $w=w_0=3$. This identifies the isotropic scaling of the turbulent state without invoking a strongly anisotropic reduction of the kinetic equation. The angular dependence is then determined by the geometry of the resonant manifold. Under a mild approximation motivated by the accumulation of energy near slow modes, we obtained the asymptotic angular spectrum
\begin{equation}
    e_\alpha=C_0 k^{-3} |\cos\theta_k|^{-1/2}= C_0 k^{-2.5} |\omega_k/\nu ^{\text{odd}}|^{-0.5},
\end{equation}
which is singular only at $\omega_\alpha\to 0$ (the slow-mode curve) and avoids the nonphysical divergence at $\theta_k\to 0$ that appears when the strongly anisotropic limit is imposed too early.

Our second main result concerns the role of helicity in determining cascade directions. Although helicity is globally sign-indefinite in three dimensions, the helical decomposition splits it into sign-definite components on each polarization branch. We showed that triads whose three legs belong to the same helicity branch behave as if constrained by a sign-definite invariant and drive an upscale transfer of energy. In contrast, mixed-helicity triads support the usual downscale transfer. As a result, even in regimes where the total energy cascade is direct, helicity induces a systematic backscatter carried by sign-definite triads. In the limiting case where one branch dominates—for example, under helicity-definite forcing—the kinetic equation admits an additional scaling associated with helicity transport, $e_\alpha\propto k^{-3.5}|\cos\theta_k|^{-1/2}$, and the energy transfer becomes predominantly inverse, mirroring the familiar role of sign-definite invariants in isotropic turbulence.

We substantiated these analytical conclusions by direct numerical evaluation of the collision integral in the standard strongly anisotropic limit. In that setting, we observed (i) a one-parameter family of zeros of the normalized angular collision integral containing the turbulent solution $(w_0,f_0)=(3,1/2)$, and (ii) a robust decomposition of the energy flux into helicity classes, with same-branch triads contributing a negative (upscale) flux and mixed-branch triads contributing a positive (downscale) flux.

Several directions follow naturally from this work. First, the slow-mode singularity highlights the limits of the diagonal, Gaussian closure underlying weak turbulence theory: arbitrarily small frequencies enter the collision integral as the system size grows, suggesting that a refined kinetic description incorporating near-resonant interactions and off-diagonal correlations may be required to fully describe the interface between waves and slow modes. 

An alternative approach is to introduce a physical regularization that separates the wave manifold from the slow modes. One possible mechanism is the addition of weak stratification, which lifts the degeneracy at zero frequency. This is analogous to the treatment of internal inertia–gravity waves in strongly stratified flows, where a small rotation rate was introduced to regularize the kinetic equation and separate slow and wave dynamics \cite{shavit2026wave}.

A second direction is to investigate the Casimir structure associated with helicity in these systems. Understanding how helicity appears as a Casimir of the underlying fluid dynamics may reveal additional constraints on the turbulent state beyond those captured by the kinetic equation, and could clarify how sign-indefinite invariants organize cascade directions in anisotropic wave turbulence.

Finally, although odd waves and inertial waves share the same turbulent spectrum within the framework developed here, their interaction structures and symmetry properties differ. A systematic comparison of the two systems may therefore reveal dynamical regimes where their cascade behavior diverges, shedding further light on the role of parity breaking and non-Hermitian wave dynamics in turbulence. This direction would naturally extend the comparison between rotating fluids and odd-viscous hydrodynamics initiated in \cite{de2024pattern} to the regime of wave turbulence.

\textit{\textbf{Acknowledgments}}.
We thank Jalal Shatah, Gregory Falkovich, and Anna Frishman for many fruitful discussions.
%\begin{acknowledgments}
%We thank Gregory Falkovich and Vincent Labarre for useful discussions. 
This work was supported by the Simons Foundation and the Simons Collaboration on Wave Turbulence.  S.~A.~S. is supported by the CHE/PBC Fellowship.

\section{Appendix} 
The code used to produce the numerical results of this work is openly available at \url{https://github.com/michalshavitNYU/anisotropic-wave-turbulence}.

\subsection{The non-canonical Hamiltonian structure}
To evaluate $\delta E / \delta\bm{\omega}$, note that
\begin{equation}
\delta E = \int \mathbf{v}\cdot\delta\mathbf{v}\,d\mathbf{x}
= \int (\nabla\times\bm{\psi})\cdot\delta\mathbf{v}\,d\mathbf{x}
= \int \bm{\psi}\cdot\delta\bm{\omega}\,d\mathbf{x},
\end{equation}
where $\bm{\psi}$ is a divergence-free vector potential defined by $\nabla\times\bm{\psi} = \mathbf{v}$. Substituting into \eqref{eq:hamil} recovers the Euler equation \eqref{eq:euler} exactly.
The helicity $H$ is a \emph{Casimir} of the Poisson structure \eqref{eq:poisson}, it is conserved because the velocity $\mathbf{v}$ lies in the kernel of $\mathcal{J}$. Indeed,
\begin{align}
\partial_{t}H
&= 2\int \mathbf{v}\cdot\partial_{t}\bm{\omega}\,d\mathbf{x}
= 2\int \mathbf{v}\cdot\mathcal{J}\,\frac{\delta E}{\delta\bm{\omega}}\,d\mathbf{x} \nonumber\\
&= -\,2\int \mathcal{J}\mathbf{v}\cdot\frac{\delta E}{\delta\bm{\omega}}\,d\mathbf{x} = 0,
\end{align}
where skew-symmetry of $\mathcal{J}$ and $\mathcal{J}\mathbf{v}=0$ imply the result.

\addSAS{\subsection{Triangle areas in two and three dimensions}}
In this appendix we evaluate numerically the ratio
\begin{align}
    R(\theta_\alpha) \equiv \frac{\obs{A_{3D}}_{\beta \gamma}}{\obs{A_{2D}}_{\beta \gamma}},
\end{align}
introduced in section "Validity of the projected-area approximation". The goal is to quantify the error introduced by replacing the projected triangle area by the full three-dimensional area.

\addSAS{\subsubsection{Analytical result for an isotropic spectrum}}
\label{app:isotropic-area-ratio}

Before evaluating the anisotropic case numerically, it is useful to compute the corresponding area ratio analytically for an isotropic distribution. Let
\begin{align}
    \widehat{\mathbf n}
    \equiv
    \frac{\mathbf{k}_\alpha\times\mathbf{k}_\beta}
    {\abs{\mathbf{k}_\alpha\times\mathbf{k}_\beta}}
\end{align}
denote the unit normal to the triad plane. %and let $\psi$ be the angle betweenc$\widehat{\mathbf n}$ and the vertical direction $\widehat{\mathbf z}$. 
Since
the two-dimensional triangle is the horizontal projection of the
three-dimensional triangle,
\begin{align}
    A_{2D}
    =
    A_{3D}\abs{\widehat{\mathbf n}\cdot\widehat{\mathbf z}}
    \label{eq:projected-area-normal}
\end{align}

\begin{comment}
\begin{align}
    A_{2D}
   =
    A_{3D}\abs{\widehat{\mathbf n}\cdot\widehat{\mathbf z}}
    =
    A_{3D}\abs{\cos\psi}.
    \label{eq:projected-area-normal}
\end{align}
\end{comment}

We first keep $\mathbf{k}_\alpha$ fixed and average over the orientations of
the partner wavevectors. Let $\theta_\alpha$ denote the polar angle between
$\mathbf{k}_\alpha$ and $\widehat{\mathbf z}$. Because the spectrum and the
integration measure are isotropic, the normal $\widehat{\mathbf n}$ is
uniformly distributed in the plane perpendicular to $\mathbf{k}_\alpha$.
Introducing an orthonormal basis
$\{\widehat{\mathbf e}_{\theta_\alpha},
\widehat{\mathbf e}_{\phi_\alpha}\}$ in this plane, we may write
\begin{align}
    \widehat{\mathbf n}
    =
    \cos\varphi\,
    \widehat{\mathbf e}_{\theta_\alpha}
    +
    \sin\varphi\,
    \widehat{\mathbf e}_{\phi_\alpha},
    \qquad
    0\leq\varphi<2\pi.
\end{align}
Since
\begin{align}
    \widehat{\mathbf e}_{\theta_\alpha}\cdot\widehat{\mathbf z}
    =
    -\sin\theta_\alpha,
    \qquad
    \widehat{\mathbf e}_{\phi_\alpha}\cdot\widehat{\mathbf z}
    =
    0,
\end{align}
the projection factor becomes
\begin{align}
    \abs{\widehat{\mathbf n}\cdot\widehat{\mathbf z}}
    =
    \sin\theta_\alpha\,\abs{\cos\varphi}.
\end{align}
Its orientational average is therefore
\begin{align}
    \left\langle
    \abs{\widehat{\mathbf n}\cdot\widehat{\mathbf z}}
    \right\rangle_{\varphi}
    &=
    \frac{1}{2\pi}
    \int_0^{2\pi}
    \sin\theta_\alpha\,\abs{\cos\varphi}\,\dd\varphi
    \nonumber\\
    &=
    \frac{2}{\pi}\sin\theta_\alpha.
    \label{eq:isotropic-projection-factor}
\end{align}
For an isotropic spectrum, the integrand is independent of the orientation $\varphi$ except through the projection factor in Eq.~\eqref{eq:projected-area-normal}, so that the orientational average factorizes:
\begin{align}
    \obs{A_{2D}}_{\beta\gamma}
    =
    \frac{2}{\pi}\sin\theta_\alpha\,
    \obs{A_{3D}}_{\beta\gamma},
\end{align}
and hence
\begin{align}
    R_{\mathrm{iso}}(\theta_\alpha)
    \equiv
    \frac{\obs{A_{3D}}_{\beta\gamma}}
         {\obs{A_{2D}}_{\beta\gamma}}
    =
    \frac{\pi}{2\sin\theta_\alpha}.
    \label{eq:isotropic-fixed-alpha-ratio}
\end{align}
In particular, on the slow manifold $k_\alpha^z\to0$, for which
$\theta_\alpha\to\pi/2$,
\begin{align}
    R_{\mathrm{iso}}(\theta_\alpha)
    \xrightarrow[k_\alpha^z\to0]{}
    \frac{\pi}{2}
    \simeq 1.57.
\end{align}
Thus, even when the distinguished leg lies in the horizontal plane, an
isotropic distribution leaves the partner legs at generic inclinations, and
the projected and three-dimensional areas do not coincide.

We may further average over the orientation of $\mathbf{k}_\alpha$. For an isotropic distribution, $\widehat{\mathbf k}_\alpha$ is uniformly distributed on the unit sphere, with normalized measure
\begin{align}
\frac{\dd\Omega_\alpha}{4\pi}
=
\frac{1}{4\pi}
\sin\theta_\alpha \dd\theta_\alpha \dd\phi_\alpha.
\end{align}
Using Eq.~\eqref{eq:isotropic-projection-factor}, we therefore obtain
\begin{align}
\left\langle
\abs{\widehat{\mathbf n}\cdot\widehat{\mathbf z}}
\right\rangle
&=
\frac{1}{4\pi}
\int_0^{2\pi}\dd\phi_\alpha
\int_0^\pi \sin\theta_\alpha\dd\theta_\alpha
\left\langle
\abs{\widehat{\mathbf n}\cdot\widehat{\mathbf z}}
\right\rangle_{\varphi}
\nonumber\
&=
\frac12.
\label{eq:isotropic-spherical-projection}
\end{align}
So that the ratio of the fully integrated areas is

\begin{align}
    \frac{\obs{A_{3D}}}{\obs{A_{2D}}}
    =
    2.
\end{align}

%\label{eq:isotropic-integrated-area-ratio}

This value is independent of the radial form of the isotropic spectrum: it
follows solely from rotational invariance and from the mean projection
$\left\langle
\abs{\widehat{\mathbf n}\cdot\widehat{\mathbf z}}
\right\rangle=1/2$ of a uniformly oriented surface element.
The smaller ratio obtained for the odd-wave spectrum therefore directly
quantifies the concentration of the triad measure toward horizontally
oriented configurations.

\addSAS{\subsubsection{Numerical computation for the turbulent anisotropic spectrum}}
 We now describe the numerical evaluation of the averaged areas. The average over the triangle area in N-dimensions is given by:
\begin{align}
    \obs{A_{ND}}_{\beta \gamma} &= \int \dd V_\beta \dd V_\gamma A_{ND} e_\beta e_\gamma \delta \leri{\mathbf{k}_\alpha+\mathbf{k}_\beta+\mathbf{k}_\gamma } \nonumber \\
    % &= \frac{C_0^2}{2} \int \rho_\beta \dd \rho_\beta \int \dd k_\beta^z  \int \rho_\gamma \dd \rho_\gamma \int \dd k_\gamma^z \abs{k_\beta^z}^{-\frac{1}{2}} \abs{k_\gamma^z}^{-\frac{1}{2}} \nonumber \\
    % & \leri{\rho_\beta^2 + {k_\beta^z}^2}^{-\frac{5}{4}} \leri{\rho_\gamma^2 + {k_\gamma^z}^2}^{-\frac{5}{4}} \frac{A_{ND}}{A_{2D}} \delta \leri{k_\alpha^z +k_\beta^z +k_\gamma^z } \,,
    &= \frac{C_0^2}{2} \int \rho_\beta \dd \rho_\beta \int \dd k_\beta^z  \int \rho_\gamma \dd \rho_\gamma \abs{k_\beta^z}^{-\frac{1}{2}} \abs{k_\alpha^z +k_\beta^z}^{-\frac{1}{2}} \nonumber \\
    & \leri{\rho_\beta^2 + {k_\beta^z}^2}^{-\frac{5}{4}} \leri{\rho_\gamma^2 + \leri{k_\alpha^z +k_\beta^z}^2}^{-\frac{5}{4}} \frac{A_{ND}}{A_{2D}}  \,.
\end{align}
%As suggested by~\cite{labarre2024kinetics,lvov2012resonant}, 
We will compute the integration limits within the kinematic box. It is then convenient to perform the following change of variables:
\begin{align}
    \rho_\alpha &< \rho_\beta + \rho_\gamma \longrightarrow& u&\equiv \frac{\rho_\beta+\rho_\gamma}{2\rho_\alpha}> \frac{1}{2}\,, \label{eqn:kinematicBox1}\\
    \rho_\beta &< \rho_\alpha + \rho_\gamma \longrightarrow& v&\equiv \frac{\rho_\beta-\rho_\gamma}{2\rho_\alpha}< \frac{1}{2}\,,\label{eqn:kinematicBox2} \\
    \rho_\gamma &< \rho_\alpha + \rho_\beta \longrightarrow& v&\equiv \frac{\rho_\beta-\rho_\gamma}{2\rho_\alpha}> -\frac{1}{2}\,, \label{eqn:kinematicBox3}
\end{align}
To avoid numerical instabilities, we emphasize an exclusion of three areas: First, $u\approx\frac{1}{2}$, and $v\approx\leri{\pm \frac{1}{2}}$, which correspond to zero area, that is, they bear zero weight, and are therefore safely excluded. Numerically, we add the regulators $h_u, h,d$ as following:
\begin{align} \label{eq:areaRatio}
    \obs{A_{ND}}_{\beta \gamma} &= C_0^2 \rho_\alpha^4 \int_{\frac{1}{2}+10^{-d}}^{hu} \dd u \int_{-\frac{1}{2}+10^{-d}}^{\frac{1}{2}-10^{-d}} \dd v \int \dd k_\beta^z  \frac{A_{ND}}{A_{2D}} \nonumber \\
    & \abs{k_\beta^z}^{-\frac{1}{2}} \abs{k_\alpha^z +k_\beta^z}^{-\frac{1}{2}} \leri{\rho_\alpha^2 \leri{u+v}^2 + {k_\beta^z}^2}^{-\frac{5}{4}}  \nonumber \\
    & \leri{\rho_\alpha^2 \leri{u-v}^2 + \leri{k_\alpha^z +k_\beta^z}^2}^{-\frac{5}{4}} \leri{u+v} \leri{u-v} \,, \\
    \int \dd k_\beta^z &=  \left\{ \begin{array}{cl}
        \text{if } k_\alpha^z <0: & \int_{-10^h}^{-10^{-d}} \dd k_\beta^z + \int_{10^{-d}}^{-k_\alpha^z-10^{-d}} \dd k_\beta^z  \\
        & + \int_{-k_\alpha^z+10^{-d}}^{10^h} \dd k_\beta^z \\
        \text{if } k_\alpha^z > 0:& \int_{-10^h}^{-k_\alpha^z-10^{-d}} \dd k_\beta^z + \int_{-k_\alpha^z+10^{-d}}^{-10^{-d}} \dd k_\beta^z \\
        & + \int_{10^{-d}}^{10^h} \dd k_\beta^z \\
    \end{array} \right. 
\end{align}
Note that, in addition to regulating the kinematic box, as explained above, we also regulate around $k_\beta^z = 0$ and $k_\beta^z = -k_\alpha^z $, which have a branch point singularity. The triangle areas are accordingly given by
\begin{align}
    A_{2D} &=\frac{\rho_\alpha^2}{4} \sqrt{4(u+v)^2 (u-v)^2 - \leri{2u^2 + 2v^2 -1}^2} \nonumber \\
    A_{3D} &= \frac{1}{4} \text{sqrt}\left[4 \leri{\rho_\alpha^2 (u+v)^2+ {k_\beta^z}^2} \right. \nonumber \\
    & \left. \leri{\rho_\alpha^2 (u-v)^2+ \leri{k_\alpha^z + k_\beta^z}^2} - \right. \nonumber \\
    & \left. \leri{\rho_\alpha^2 (2u^2+2v^2-1)+ {k_\beta^z}^2 + \leri{k_\alpha^z+k_\beta^z}^2 -{k_\alpha^z}^2 }^2 \right] \,.
\end{align}
The numerical results are presented in Fig.~\ref{fig:triangleAreas} and Fig.~\ref{fig:arearatio}, the numerical parameters are: $hu=40,\ h=d=8$, and absolute tolerance given by 1e-5. Fig.~\ref{fig:arearatio} was generated by Claude (Anthropic) by extending our code for Fig.~\ref{fig:triangleAreas}; the code is available in our GitHub repository.

\begin{figure}[t]
    \centering
    \includegraphics[width=\columnwidth]{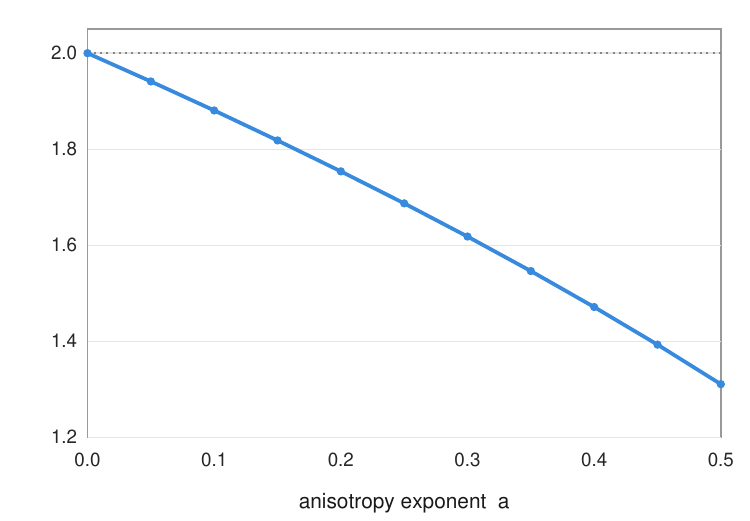}
    \caption{\label{fig:arearatio}
    Ratio of the averaged triangle areas $\obs{A_{3D}}/\obs{A_{2D}}$ as a
    function of the angular exponent $a$ for the spectral family
    $e_k\propto k^{-3}|\cos\theta_k|^{-a}$. The ratio decreases monotonically
    from the exact isotropic value $2$ (at $a=0$, dotted line) to
    $\simeq 1.31$ for the odd-wave spectrum $a=\tfrac12$, showing that the
    projected-area approximation is increasingly accurate as the measure
    concentrates on the slow-mode manifold.}
\end{figure}

\section{Parametrization of the resonant manifold}
Consider the collision integral of a triad of waves forming a triangle, as given in Eq.~\eqref{eq:kinsimple}, with the energy density given by Eq.~\eqref{eq:turbspec}. This homogeneous solution makes the collision integral separable, so the angular part decouples from the rest of the integration and can be solved separately, as shown below.\\

We work in cylindrical coordinates $\mathbf{k}_\alpha = \leri{\rho_\alpha, \phi_\alpha, k_\alpha^z}$, and express the volume integration in terms of $\leri{k_\alpha \equiv \sqrt{\rho_\alpha^2 + {k_\alpha^z}^2}, \phi_\alpha, k_\alpha^z}$:
    \begin{align}
        % K &= +\sqrt{\rho_\alpha^2 + {k_z}^2} \longrightarrow \rho_\alpha = +\sqrt{K^2 - k_z^2} \,, \nonumber \\
        % \dd \rho_\alpha &= \frac{\partial \rho_\alpha}{\partial K} \dd K = \frac{K}{\sqrt{K^2 - k_z^2}} \dd K = \frac{K}{\rho_\alpha} \dd K \,. \nonumber \\ 
        \int \dd V_\alpha &= \int \rho_\alpha \dd \rho_\alpha  \int \dd \phi_\alpha \int \dd k_\alpha^z = \int k_\alpha \dd k_\alpha  \int \dd \phi_\alpha \int \dd k_\alpha^z\,.
    \end{align}
The collision integral then reads:
\begin{align}
    \partial_t e_k &= \sum_{\beta,\gamma}\iint k_\beta \dd k_\beta \dd k_\beta^z \iint k_\gamma \dd k_\gamma \dd k_\gamma^z k_\alpha^z \left|\Gamma_{\alpha\beta\gamma}\right|^{2} \nonumber \\
    &\left(k_\alpha^z e_{\beta}e_{\gamma}+k_\beta^z e_{\alpha}e_{\gamma}+k_\gamma^z e_{\alpha}e_{\mathbf{\beta}}\right) \delta\left(\omega_{\alpha}+\omega_{\beta}+\omega_{\gamma}\right)  \nonumber \\
    &\delta \left(k_\alpha^z + k_\beta^z + k_\gamma^z \right) I_\phi\,, \label{eq:kineticEqn2}
\end{align}
where we define
\begin{align}
    I_\phi &\equiv \iint \dd \phi_\beta \dd \phi_\gamma \delta\left( \rho_\alpha \cos(\phi_\alpha) + \rho_\beta \cos(\phi_\beta) + \rho_\gamma \cos(\phi_\gamma)\right) \nonumber \\
    &\delta \left(\rho_\alpha \sin(\phi_\alpha) + \rho_\beta \sin(\phi_\beta) + \rho_\gamma \sin(\phi_\gamma)\right)
\end{align}
As mentioned, the angular integral, $I_\phi$, can be solved separately by the following change of variables:
\begin{align}
    s &=  \rho_\alpha \cos(\phi_\alpha) + \rho_\beta \cos(\phi_\beta) + \rho_\gamma \cos(\phi_\gamma)\,, \\
        % v 
        t &= \rho_\alpha \sin(\phi_\alpha) + \rho_\beta \sin(\phi_\beta) + \rho_\gamma \sin(\phi_\gamma)\,.
\end{align}
Substituting back, with the Jacobian, the integral reads:
\begin{align}
    I_\phi &= \iint \frac{1}{\rho_\beta \rho_\gamma \sin \leri{\phi_\beta-\phi_\gamma}} \delta(s) \delta(t) \dd s \dd t\,.
\end{align}
To perform the integration, we rewrite the sine as a function of $(s, t)$ and set them to zero:
	\begin{align}
		\left. \rho_\beta \rho_\gamma \sin(\phi_\beta - \phi_\gamma) \right|_{s=t=0} &= \pm \sqrt{\rho_\beta^2 \rho_\gamma^2-\frac{1}{4}\leri{\rho_\beta^2 +\rho_\gamma^2 - \rho_\alpha^2}^2} \nonumber \\
        &\equiv \pm 2A_{2D} \,.
	\end{align}
We note that $A_{2D}$ corresponds to the area enclosed by the triangle of $\rho_\alpha$, $\rho_\beta$ and $\rho_\gamma$ via Heron's formula. Alternatively, one can think of the area enclosed by the triangle of three vectors $\bm{k}_\alpha\,, \bm{k}_\beta$ and $\bm{k}_\gamma$, projected on the $XY$ plane.\\
When we consider the geometrical interpretation of the above equation, it is clear that the negative branch does not participate: $\sin(\phi_\beta - \phi_\gamma)$ is the sine of an angle in a triangle; therefore, it must be positive. Finally:
    \begin{align}
        I_\phi &= \frac{1}{2A_{2D}} \,.
    \end{align}
We return to Eqn.~\eqref{eq:kineticEqn2}, and use the dispersion relation calculated in Eq.~\eqref{eq:dispersion}:
\begin{align}
    \delta\left(\omega_{\alpha}+\omega_{\beta}+\omega_{\gamma}\right) &=  \frac{\delta\leri{\alpha k_\alpha^z k_\alpha + \beta k_\beta^z k_\beta +\gamma k_\gamma^z k_\gamma}}{\abs{\nu_\text{odd}}} \,.
\end{align}
Lastly, the interaction term is given by~\cite{waleffe1992nature}:
\begin{align}
    \left|\Gamma_{\alpha\beta\gamma}\right|^{2} &=  \frac{A^2_{3D}}{4 {k_\alpha^z}^2 k_\alpha^2 k_\beta^2 k_\gamma^2} \leri{\beta k_\beta - \gamma k_\gamma}^2 \leri{\alpha k_\alpha + \beta k_\beta + \gamma k_\gamma}^2\,, \\
     A_{3D} &\equiv + \frac{1}{4} \sqrt{4 k_\beta^2 k_\gamma^2-\leri{k_\beta^2 +k_\gamma^2 - k_\alpha^2}^2} \,.
\end{align}
Using the bi-homogeneous spectrum~\eqref{eq:turbspec} with $\cos(\theta)=\frac{k_\alpha^z}{k_\alpha}$, the sum over interaction coefficients reads
\begin{align}
    k_{z}e_{\beta}e_{\gamma}+&p_{z} e_{\alpha}e_{\gamma}+q_{z}e_{\alpha}e_{\mathbf{\beta}} = C_0^2 \left( k_\alpha^z \leri{k_\beta k_\gamma}^{-w+f} \abs{k_\beta^z k_\gamma^z}^{-f} \right. \nonumber \\
    &\left. +k_\beta^z \leri{k_\alpha k_\gamma}^{-w+f} \abs{k_\alpha^z k_\gamma^z}^{-f} + \dots \right. \nonumber \\
    & \left. k_\gamma^z \leri{k_\alpha k_\beta}^{-w+f} \abs{k_\alpha^z k_\beta^z}^{-f} \right) \,.
\end{align}
We can now substitute all of the above and perform the integration over $k_\gamma^z$ using the delta function:
\begin{align}
    \partial_t e_k &= \sum_{\beta,\gamma} \int \dd k_\beta  \int \dd k_\gamma \int\limits_{-k_\beta}^{k_\beta} \dd k_\beta^z \frac{ C_0^2 A^2_{3D}}{8 A_{2D} k_\alpha^z k_\alpha^2 k_\beta k_\gamma} \nonumber \\
    & \leri{\beta k_\beta - \gamma k_\gamma}^2 \leri{\alpha k_\alpha + \beta k_\beta + \gamma k_\gamma}^2 \times \dots \nonumber \\
    & \frac{\delta \leri{\alpha k_\alpha^z k_\alpha + \beta k_\beta^z k_\beta -\gamma \left(k_\alpha^z + k_\beta^z \right) k_\gamma}}{\abs{\nu_\text{odd}}} \nonumber \\
    &\left( k_\alpha^z \leri{k_\beta k_\gamma}^{-w+f} \abs{k_\beta^z \left(k_\alpha^z + k_\beta^z \right)}^{-f} + \right. \nonumber \\
    &\left. k_\beta^z \leri{k_\alpha k_\gamma}^{-w+f} \abs{k_\alpha^z \left(k_\alpha^z + k_\beta^z \right)}^{-f} - \right. \nonumber \\
    & \left. \left(k_\alpha^z + k_\beta^z \right)\leri{k_\alpha k_\beta}^{-w+f} \abs{k_\alpha^z k_\beta^z}^{-f} \right)\,.
\end{align}
We note that, in general, since $k_\gamma^z$ is bounded by $k_\gamma$ via $\abs{k_\gamma^z} \leq k_\gamma$, we must ensure $k_\beta^z$ obeys:
\begin{align} \label{eq:kBetaZ_condition}
    \abs{k_\alpha^z + k_\beta^z } &\leq k_\gamma \nonumber \\
     \max \{-k_\beta, -k_\gamma-k_\alpha^z\} &\leq k_\beta^z \leq \min \{k_\beta, k_\gamma-k_\alpha^z\} \,.
\end{align}

\section{Evaluating the collision integral, in the very anisotropic approximation}

Proceeding, we assume the very \textbf{anisotropic limit} where the projection of each wave-vector on the $z$-direction is much smaller than the absolute value of the wave: $k_\alpha \gg k_\alpha^z\,, k_\beta \gg k_\beta^z\,, k_\gamma \gg k_\gamma^z$. As a result, the condition described in Eq.~\eqref{eq:kBetaZ_condition} is trivially satisfied. Moreover, the 2D area is approximated by the 3D area, further simplifying the calculation.\\

Using properties of the delta function:
\begin{align} \label{eq:pzDeltaFunc}
    \delta \leri{\alpha k_\alpha^z k_\alpha + \beta k_\beta^z k_\beta -\gamma \leri{k_\alpha^z+k_\beta^z} k_\gamma} &= \frac{\delta \leri{k_\beta^z - k_\alpha^z \frac{\gamma k_\gamma - \alpha k_\alpha}{\beta k_\beta - \gamma k_\gamma}}}{\abs{\beta k_\beta - \gamma k_\gamma}}\,.
\end{align}
In the special case where $k_\beta = k_\gamma$ and $\beta = \gamma$, the delta function is proportional to $\delta\leri{k_\alpha^z \leri{\alpha k_\alpha - \beta k_\beta}}$ requiring that either $k_\alpha^z$ vanishes (see discussion on slow modes in the main text), or an equilateral triangle in \ac{3D} $k_\alpha=k_\beta=k_\gamma$, on the branch: $\alpha=\beta=\gamma$.\\

Thus, the collision integral reads:
\begin{align} \label{eq:kinEqnKs}
    \partial_t e_k &= \sum_{\beta,\gamma}\int  \dd k_\beta \int \dd k_\gamma  \frac{C_0^2 A_{3D} }{8 \abs{\nu_\text{odd}} k_\alpha^z k_\alpha^2 k_\beta k_\gamma} \nonumber \\
    &\left|\beta k_\beta - \gamma k_\gamma\right| \leri{\alpha k_\alpha + \beta k_\beta + \gamma k_\gamma}^2 \left( k_\alpha^z \leri{k_\beta k_\gamma}^{-w+f}  \right. \nonumber \\
    & \left. \abs{\leri{k_\alpha^z}^2 \frac{\gamma k_\gamma - \alpha k_\alpha}{\beta k_\beta - \gamma k_\gamma} \frac{\beta k_\beta - \alpha k_\alpha}{\beta k_\beta - \gamma k_\gamma}}^{-f} + \right.\nonumber \\
    &\left. k_\alpha^z \frac{\gamma k_\gamma - \alpha k_\alpha}{\beta k_\beta - \gamma k_\gamma} \leri{k_\alpha k_\gamma}^{-w+f} \abs{\leri{k_\alpha^z}^2 \frac{\beta k_\beta - \alpha k_\alpha}{\beta k_\beta - \gamma k_\gamma}}^{-f} \right. \nonumber \\
    & \left. -k_\alpha^z \frac{\beta k_\beta - \alpha k_\alpha}{\beta k_\beta - \gamma k_\gamma}\leri{k_\alpha k_\beta}^{-w+f} \abs{\leri{k_\alpha^z}^2 \frac{\gamma k_\gamma - \alpha k_\alpha}{\beta k_\beta - \gamma k_\gamma}}^{-f} \right) \,.
\end{align} 
It is now apparent that around the singular point of an equilateral triangle, namely when the difference between each pair of $k$'s scales like $\tilde\epsilon$ and $\alpha = \beta = \gamma$, the integrand scales like $\tilde\epsilon$ and its limit should vanish. However, this introduces numerical instability, which we shall later address.\\

Additionally, it is important to note that the integration limits on the set of $k$'s must obey the triangle inequality. Following~\cite{labarre2024kinetics,lvov2012resonant}, we will compute the integration limits within the kinematic box. Moreover, it is convenient to perform the following change of variables to $(u,v)$:
\begin{align}
    k_\alpha &< k_\beta + k_\gamma \longrightarrow& u&\equiv \frac{k_\beta+k_\gamma}{2k_\alpha}> \frac{1}{2}\,, \label{eqn:kinematicBox1}\\
    k_\beta &< k_\alpha + k_\gamma \longrightarrow& v&\equiv \frac{k_\beta-k_\gamma}{2k_\alpha}< \frac{1}{2}\,,\label{eqn:kinematicBox2} \\
    k_\gamma &< k_\alpha + k_\beta \longrightarrow& v&\equiv \frac{k_\beta-k_\gamma}{2k_\alpha}> -\frac{1}{2}\,, \label{eqn:kinematicBox3}
\end{align}
Accordingly, the integral transforms as:
\begin{align}
    \int \dd k_\beta \int \dd k_\gamma &= 2 k_\alpha^2 \int\limits_{\frac{1}{2}}^{\infty} \dd u \int\limits_{-\frac{1}{2}}^{\frac{1}{2}} \dd v \,.
\end{align}
In the new parameters $(u,v)$, the kinetic equation in the very anisotropic limit reads:
\begin{align} \label{eq:Collision_uv}
    \partial_t e_k &= \sum_{\beta,\gamma}\int\limits_{\frac{1}{2}}^{\infty} \dd u \int\limits_{-\frac{1}{2}}^{\frac{1}{2}} \dd v \frac{C_0^2 \sqrt{\leri{4u^2 -1} \leri{1-4v^2}} }{2^4 \abs{\nu_\text{odd}}\leri{u^2-v^2}} \nonumber \\
    &\left|u (\beta-\gamma) +v (\beta+\gamma) \right| \leri{\alpha + u (\beta+\gamma) + v (\beta-\gamma)}^2 \nonumber \\
    & k_\alpha^{3-2w+2f} \left|k_\alpha^z\right|^{-2f} \left( \leri{u^2-v^2}^{-w+f} \right. \nonumber \\
    & \left. \abs{\frac{\leri{\alpha -\beta (u+v)}\leri{\alpha -\gamma (u-v)}}{\leri{u(\beta -\gamma)+v(\beta+\gamma)}^2}}^{-f} - \right. \nonumber \\
    & \left. \frac{\alpha-\gamma(u-v)}{u(\beta-\gamma)+v(\beta +\gamma)}\leri{u-v}^{-w+f} \right. \nonumber \\
    & \left. \abs{\frac{\alpha-\beta(u+v)}{u(\beta-\gamma)+v(\beta +\gamma)}}^{-f}+\frac{\alpha-\beta(u+v)}{u(\beta-\gamma)+v(\beta +\gamma)} \right. \nonumber \\
    & \left. \leri{u+v}^{-w+f} \abs{\frac{\alpha-\gamma(u-v)}{u(\beta-\gamma)+v(\beta +\gamma)} }^{-f} \right)\,.
\end{align}
Accordingly, the radial flux is given, using Eq.~\eqref{eq:fluxang}, by
\begin{align}
\Pi^r_\alpha(w_0) &=\left. \frac{C_0^2}{2k^{d-1}}\frac{d}{dw}St_{\Omega}\right|_{w=w_{0}} \\
&= \left. \frac{1}{2k^{d-1}}\frac{d}{dw} \leri{\frac{1}{k^{d-2w}} St_{\alpha} }\right|_{w=w_{0}} \nonumber \\
&= -\leri{k_\alpha}^{2f-(d-1)} \left|k_\alpha^z\right|^{-2f} \sum_{\beta,\gamma}\int\limits_{\frac{1}{2}}^{\infty} \dd u \int\limits_{-\frac{1}{2}}^{\frac{1}{2}} \dd v \nonumber \\
& \frac{C_0^2 \sqrt{\leri{4u^2 -1} \leri{1-4v^2}} }{2^5 \abs{\nu_\text{odd}}\leri{u^2-v^2}} \left|u (\beta-\gamma) +v (\beta+\gamma) \right| \nonumber \\
& \leri{\alpha + u (\beta+\gamma) + v (\beta-\gamma)}^2 \left( \leri{u^2-v^2}^{-w_0+f}  \right. \nonumber \\
    & \left. \ln \leri{u^2-v^2}  \abs{\frac{\leri{\alpha -\beta (u+v)}\leri{\alpha -\gamma (u-v)}}{\leri{u(\beta -\gamma)+v(\beta+\gamma)}^2}}^{-f} - \right. \nonumber \\
    & \left. \frac{\alpha-\gamma(u-v)}{u(\beta-\gamma)+v(\beta +\gamma)}\leri{u-v}^{-w_0+f} \ln \leri{u-v} \right. \nonumber \\
    &  \left. \abs{\frac{\alpha-\beta(u+v)}{u(\beta-\gamma)+v(\beta +\gamma)}}^{-f}+ \frac{\alpha-\beta(u+v)}{u(\beta-\gamma)+v(\beta +\gamma)} \right. \nonumber \\
    & \left. \leri{u+v}^{-w_0+f} \ln \leri{u+v} \abs{\frac{\alpha-\gamma(u-v)}{u(\beta-\gamma)+v(\beta +\gamma)} }^{-f} \right) \,.
\end{align}

\begin{comment}
Old flux calculcation:
    \begin{align}
    \Pi_\alpha^r &=\left. \frac{1}{2k_\alpha^2} \frac{\dd \leri{\partial_t e_\alpha}}{\dd w} \right|_{w=w_0} \nonumber \\
    &= - \frac{C_0^2}{2^5 \abs{\nu_\text{odd}}} \sum_{\beta,\gamma}\int\limits_{\frac{1}{2}}^{\infty} \dd u \int\limits_{-\frac{1}{2}}^{\frac{1}{2}} \dd v \frac{ \sqrt{\leri{4u^2 -1} \leri{1-4v^2}} }{\leri{u^2-v^2}} \nonumber \\
    &\left|u (\beta-\gamma) +v (\beta+\gamma) \right| \leri{\alpha + u (\beta+\gamma) + v (\beta-\gamma)}^2 \nonumber \\
    & k_\alpha^{1-2w_0+4f} \abs{k_\alpha^z}^{-4f} \left(\leri{u^2-v^2}^{-w_0+2f} \ln\leri{k_\alpha^2 \leri{u^2-v^2}} \right. \nonumber \\
    &\left. \abs{\frac{\leri{\alpha -\beta (u+v)}\leri{\alpha -\gamma (u-v)}}{\leri{u(\beta -\gamma)+v(\beta+\gamma)}^2}}^{-2f} - \right. \nonumber \\
    & \left. \frac{\alpha-\gamma(u-v)}{u(\beta-\gamma)+v(\beta +\gamma)}\leri{u-v}^{-w_0+2f} \right. \nonumber \\
    & \left. \ln\leri{k_\alpha^2 (u-v)} \abs{\frac{-\alpha+\beta(u+v)}{u(\beta-\gamma)+v(\beta +\gamma)}}^{-2f}+\right. \nonumber \\
    & \left. \frac{\alpha-\beta(u+v)}{u(\beta-\gamma)+v(\beta +\gamma)} \leri{u+v}^{-w_0+2f} \right. \nonumber \\
    & \left. \ln\leri{k_\alpha^2 (u+v)} \abs{\frac{\alpha-\gamma(u-v)}{u(\beta-\gamma)+v(\beta +\gamma)} }^{-2f} \right) \,.
\end{align}
\end{comment}

\section{Numerical evaluation} \label{app:numericalEval}
The collision integral given in Eq.~\eqref{eq:Collision_uv} can be easily evaluated using HCubature module in the Julia programming language~\cite{HCubature}: a multidimensional integration computed adaptively, by dividing the integration volume into smaller sections, until it converges.\\

To avoid numerical instabilities, we emphasize an exclusion of three areas: First, $u\approx\frac{1}{2}$, and $v\approx\leri{\pm \frac{1}{2}}$, which correspond to zero area, that is, they bear zero weight, and are therefore safely excluded. Moreover, assuming $f$ is positive, NaN results are obtained for the following cases:
\begin{itemize}
    \item For $u-v=1,\ \gamma=\alpha$, 
    \item For $v=0,\ \beta= \gamma$, 
    \item For $u+v=1,\ \beta = \alpha$.
    \end{itemize}
All of these result from an equilateral triangle, as can be seen from the delta function~\eqref{eq:pzDeltaFunc}. Although these points introduce a numerical instability, their analytical limit vanishes, as explained below Eq.~\eqref{eq:kinEqnKs}.\\

Numerically, we introduce two regulators, $h$ and $d$, such that:
\begin{itemize}
    \item We replace $v \in \leri{\frac{1}{2}, \infty}$ by $u \in [\frac{1}{2} + 10^{-d}, 10^h]$. 
    \item We replace $u \in \leri{-\frac{1}{2}, \frac{1}{2}}$ by
    \begin{itemize}
        \item When $\beta \neq \gamma$: $v\in \left[-\frac{1}{2}+10^{-d}, \frac{1}{2}-10^{-d} \right]$
        \item When $\beta = \gamma$: $v\in \left[-\frac{1}{2}+10^{-d}, -10^{-d}\right]$ $ \&  \left[10^{-d}, \frac{1}{2}-10^{-d} \right]$.
    \end{itemize}
\end{itemize}
The original limits are recovered at the limit $(h,d) \to \infty$, however, \addSAS{both the kinetic equation and fluxes are stable across a range of these regulatory values. The values of the regulators, $h$ and $d$, were chosen well after the numerical results had saturated. Specifically, in the figures presented in the paper, $h=6$ and $d=5$.}

To find the steady states of $e_k(w, f)$, we used the `find\_zero' function from the `Roots' package in Julia~\cite{Roots.jl}.

\bibliography{references.bib}

\end{document}